\documentclass[twocolumn]{aastex631}
\usepackage{bm}
\usepackage{amsmath}

\newcommand{\SL}[1]{{#1}}

\begin{document}

\title{Polar circumtriple planets and disks around misaligned hierarchical triple stars }

\author[0000-0003-2270-1310]{Stephen Lepp}

\author[0000-0003-2401-7168]{Rebecca G. Martin}

\affiliation{Nevada Center for Astrophysics, University of Nevada, Las Vegas, 4505 S. Maryland Pkwy., Las Vegas, NV 89154, USA}
\affiliation{Department of Physics and Astronomy,University of Nevada, Las Vegas, 4505 S. Maryland Pkwy., Las Vegas, NV 89154, USA}

\author[0000-0002-4636-7348]{Stephen H. Lubow}

\affiliation{Space Telescope Science Institute, 3700 San Martin Drive, Baltimore, MD 21218, USA}

\begin{abstract}

Observations of hierarchical triple star systems show that misalignments are common both between the angular momentum vector of the inner binary and the outer companion orbit, and between the outer binary orbit and a circumtriple gas disk.  With analytic methods and $n$-body simulations we explore the dynamics of circumtriple orbits around a misaligned hierarchical triple star. Circumtriple test particle orbits nodally precess either about the outer binary angular momentum vector (circulating orbits) or about a stationary inclination  that depends upon the binary properties (librating orbits). For a coplanar (or retrograde coplanar) triple star, 
the apsidal precession rate is maximal and the critical orbital radius outside of which all orbits are circulating is minimal.  Polar alignment of a circumtriple  gas disk requires nodal libration and therefore it can be more likely if there is a large misalignment between the inner and outer binary orbits. There are two values of the mutual misalignment, $i_{\rm c}$ and $180^\circ-i_{\rm c}$, for which the apsidal precession rate of the triple star is zero and polar alignment is possible at all orbital radii. For a circular inner binary orbit $i_{\rm c}=55^\circ$, and it changes with eccentricity of the inner binary while being insensitive to other triple star parameters.

\end{abstract}


\section{Introduction}
\label{sec:intro}

The formation of multiple star systems is common in star forming regions \citep{Duchene2013}.  Hierarchical triple stars are composed of an {\it inner binary} that is in an orbit with another companion, the {\it outer binary}.  In a triple star system, misalignments between the inclination of the inner and outer binary ($i_{\rm AB}>0$) may be common, particularly for wider outer binaries \citep{Tokovinin2017}.  Large mutual inclinations including counter rotating inner and outer binaries have been observed \citep{Tokovinin2016,Schaefer2016, Mitnyan2024}. For sufficiently highly inclined orbits, the triple star may undergo Kozai-Lidov (KL) oscillations \citep{Zeipel1910,Kozai1962,Lidov1962}, where the inclination and eccentricity of the inner binary are exchanged \citep[e.g.,][]{Naoz2016}. There is a peak in the observed mutual inclination distribution at around $i_{\rm AB}=40^\circ$ \citep{Borkovits2016}, evidence that KL oscillations are operating in triple star systems \citep{Kiseleva1998,Fabrycky2007}.

Misalignments between the orbital plane of a circumtriple disc \citep{Tobin2016,Bate2018} and the outer binary of the triple star may also be common, examples include GG Tauri A \citep{DiFolco2014,Keppler2020,Phuong2020} and GW Ori \citep{Bi2020,Kraus2020,Smallwood2021}.  The disk misalignment may occur as a result of turbulence in the molecular gas cloud \citep{Offner2010, Tokuda2014, Bate2012} or later  accretion  \citep{Bate2010, Bate2018}. The degree of misalignment can increase through the effects of stellar flybys \citep{Nealon2020} or bound stellar companions \citep[e.g.][]{Aly2015,Martin2017,Martin2022,Ceppi2023}.

The dynamics of circumtriple particle orbits can aid our understanding of the dynamics of circumtriple protoplanetary discs. The rings of the disc feel the same torque as the particle, but the rings can be in sufficiently good radial communication to  undergo solid body precession  \citep{Papaloizou1995,Larwoodetal1996}. Wave-like communication occurs in protoplanetary discs \citep{Papaloizou1983,Lubow2001,Nixon2016} and dissipation leads to alignment towards a stationary state, either coplanar to the binary \citep{Nixon2013,Facchini2013}, or polar to the binary \citep{Aly2015,Martin2017,Lubow2018,Zanazzi2018,Cuello2019} depending upon the type of nodal precession. Polar circumbinary discs have been observed during the gas disc phase \citep{Kennedy2019,Kenworthy2022} and the debris disc phase \citep{Kennedy2012}. Polar circumbinary planets have not yet been found, likely as a result of selection biases \citep{MartinDV2015,MartinDV2017}, but they may form as efficiently around polar binaries as coplanar binaries \citep{Childs2021,Childs2021b}.

Around a binary star system, circumbinary particle orbits can undergo two different types of nodal precession depending upon their initial inclination \citep[e.g.][]{Verrier2009,Farago2010,Doolin2011}. For low initial inclination, the particle orbit nodally precesses about the binary angular momentum vector, these are {\it circulating orbits}. However, for higher initial misalignment, the particle may be in a {\it librating orbit}, meaning that it precesses around the eccentricity vector of the binary.   In the quadrupole approximation of the binary potential, the particle dynamics around a binary do not depend upon the orbital radius of the test particle, only the timescale on which the precession occurs. 

Apsidal precession of the binary  leads to a dependence of the particle orbits on their separation to the binary \citep[e.g.][]{Lepp2022,Zanardi2023,Chen2024}.  General relativity (GR)  and tides drive prograde apsidal precession of the binary. Close to the binary star system, the particle orbits are similar to those around a binary  that does not undergo apsidal precession.
As a result the apsidal precession of the 
binary, the stationary inclination increases with the orbital radius of the particle. There is a critical radius outside of which there are no librating orbits, only circulating orbits \citep{Lepp2023}. Similarly, in a triple star system, the inner and outer binary undergo apsidal precession. The direction of the apsidal precession depends upon the mutual inclination between the inner and outer binary, $i_{AB}$.

 Previous work on the dynamics of circumtriple orbits has assumed that the inner and outer binary orbits are close to coplanar to each other \SL{  \citep[e.g.,][]{Lepp2023, Ceppi2023} }. 
In this work,  we examine the effect of a highly misaligned inner triple star on circumtriple orbits. We use analytic methods in the quadrupole approximation and $n$-body simulations.  In Section~\ref{triplestar} we first examine the dynamics of triple star systems with mutual misalignments. In Section~\ref{sec:orbits} we consider circumtriple test particle orbits. We draw our conclusions in Section~\ref{conc} and discuss implications for circumtriple discs and planets.

\section{Triple Star Dynamics}
\label{triplestar}

\begin{figure}
\includegraphics[width=\columnwidth]{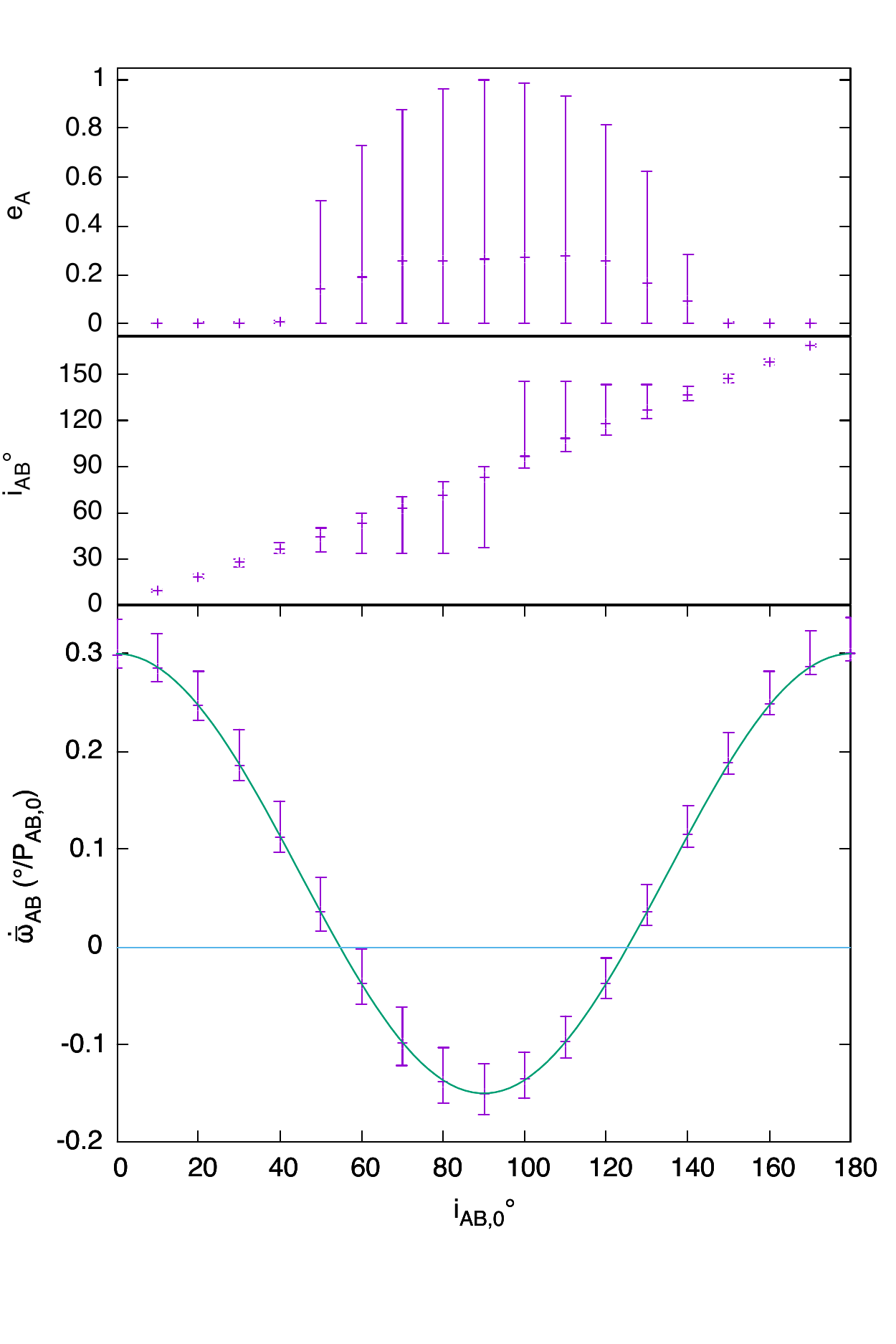}
\caption{Triple star systems with an initial mutual inclination $i_{AB,0}$ between the inner and outer binary. In each panel, $n$-body simulations are shown as points and the analytic model is plotted with a curve in the lower panel. Upper panel: The maximum eccentricity and average eccentricity of the inner binary. Middle panel: 
The  average value and range of the mutual inclination $i_{AB}$.  Lower panel: The apsidal precession rate  of the outer binary averaged over 100 initial orbital periods of the outer binary. The range shows the highest and lowest precession rate for a single outer binary orbital period. \SL{The units are degrees per initial outer binary orbital period. The green line is an analytical curve discussed in Section~\ref{sec:analytic1}.
}  } 
\label{fig:maxe}
\end{figure}

We consider a hierarchical triple star system consisting of an inner binary that is orbited by  a more distant companion.
For our standard parameters we adopt the same standard triple star parameters as \cite{Lepp2023}.
The inner binary has masses  $m_{Aa}=m_{Ab}=0.5\,\rm  M_\odot$ with a total mass of $m_A=m_{Ab}+m_{Aa}=1\,\rm M_\odot$ and orbits with semi-major axis $a_A=0.5\,\rm au$ and eccentricity $e_A=0$. We adopt the convention the $m_{Ab} < m_{Aa}$. The outer companion has mass $m_B=1 M_\odot$, semi-major axis  $a_{AB}=10\,\rm au$ and eccentricity $e_{AB}=0.5$.  The orbital period is $P_{AB}=2\pi/\sqrt{G (m_A+m_B)/a_{AB}^3}$.   The models  are scale free in mass and length so the dynamics are determined by the mass and distance ratios, which for this case they are $f_A= \frac{m_{Ab}}{m_A}=0.5$, $f_B=\frac{m_B}{m_A+m_B}=0.5$  and $a_B/a_A=20$. Note that in the general case $f_A $ ranges from 0 to $\frac{1}{2}$ while $f_B$ ranges from 0 to 1. The mass unit is the total mass of the three star system, $m_{AB}=m_A+m_B$, and the length unit is the semi-major axis of the outer binary, $a_{AB}$.

\begin{table}
\centering
\begin{tabular}{c c c }
\hline
Case & $\Omega_A$ & $\omega_A$\\
\hline\hline
A & $0^\circ$  &  $0^\circ$\\
B & $90^\circ$&  $90^\circ$ \\
C & $90^\circ$ & $0^\circ$ \\
D  &        $ 0^\circ$   &       $90^\circ$\\
\hline
\end{tabular}
\caption{The four cases that we consider for the initial  alignment of the two binaries depending  on initial values of $\Omega_A$ and $\omega_A$. }
\label{tab:cases}
\end{table}

\begin{table*}
\centering
\begin{tabular}{l c c c c c c c c c}
\hline
Configuration & $m_{Aa}$ ($M_\odot$) &$m_{Ab}$ ($M_\odot$) & $m_B$($M_\odot$) & $a_A$ (au)&$a_{AB}$(au)&$e_{A}$&$e_{AB}$&$i_{AB}^\circ$\\
\hline\hline
Standard & 0.5&0.5  &  1.0 & 0.5 & 10.0&0&0.5& 0-180&\\
Figures \ref{fig:prec2}, \ref{fig:scale2} & 0.5&0.5 & 1.0 & 0.5 & 10.0& 0.2,0.4,0.6,0.8\footnote{Note the 0.8 value only run for Fig 2 case A.}&0.5& 0-180 \\
Figure \ref{fig:scale1} ($f_A$) & 0.8, 0.9&0.2,0.1 & 1.0 & 0.5 & 10.0& 0&0.5& 0-180&\\
Figure \ref{fig:scale1} ($f_B$) & 0.5&0.5 & 0.5, 4.0 & 0.5 & 10.0& 0&0.5& 0-180&\\
Figure \ref{fig:scale1} ($a_{AB}/a_A$) & 0.5&0.5 & 1.0 & 0.5 & 5.0, 20.0 & 0&0.5& 0-180&\\
Figure \ref{fig:scale1} ($e_{AB})$ & 0.5&0.5 & 1.0 & 0.5 & 10.0& 0&0.3, 0.7& 0-180&\\
\hline
\end{tabular}
\caption{\SL{The main triple star configurations considered. Each configuration is run for all four cases of Table \ref{tab:cases}. Note that the results scale with mass and length as discussed in text.}}
\label{tab:config}
\end{table*} 

The relative alignment of the inner and outer binaries is important to the rate at which the outer binary undergoes apsidal precession.
The initial values for the orbital parameters of the inner binary, the longitude of the ascending node  $\Omega_A$ and the argument of the periapsis  $\omega_A$ are shown in Table \ref{tab:cases}.
In all cases, the outer binary initially has longitude of the ascending node $\Omega_{AB}$ and the argument of the periapsis $\omega_{AB}$ both equal to zero.  The outer binary always starts with its angular momentum vector in the $\hat z$ direction and its eccentricity vector in the $\hat x$ direction.

The mutual inclination between the orbit of the inner binary and the outer companion is $i_{AB}$. In \cite{Lepp2023}, we considered only low mutual inclinations of triple star.  In this paper,  we extend the parameter space to higher initial mutual inclinations. For sufficiently high mutual inclination the inner binary undergoes KL oscillations that exchange inclination and eccentricity.   
\SL{The configurations run for the various cases are shown in 
Table \ref{tab:config}.}

\subsection[N-body simulations]{$N$-body simulations}

Simulations in this paper made use of the {\sc rebound} $n$-body code \citep{rebound}. The simulations were integrated using the WHFast, a symplectic Wisdom-Holman integrator \citep{reboundwhfast,wh} for speed but selected simulations were run with the  IAS15, a 15th order Gauss-Radau integrator \citep{reboundias15} to verify the results.

The upper panel of Fig.~\ref{fig:maxe} shows  the maximum eccentricity  and the average eccentricity for the inner binary
as a function of the initial mutual inclination of the inner and outer binary. The initial eccentricity of the inner binary is 0 in these cases.
For our standard parameters with initial mutual inclinations between about $i_{\rm AB}=40^\circ$ and $150^\circ$, the inner binary undergoes KL oscillations. The eccentricity of the inner binary, $e_A$, and the mutual inclination, $i_{AB}$, oscillate with time \citep[see also][]{Hamers2021}. The average eccentricity can be significantly above the initial eccentricity of 0 but is still small compared with the maximum eccentricity of this oscillation. The middle panel shows the minimum, average and maximum mutual inclination of the inner and outer binaries.  The average inclination  is close to the initial value.

The lower panel of Fig.~\ref{fig:maxe} shows the average apsidal precession rate of the outer binary as a function of the initial mutual inclination. \SL{ To numerically estimate the precession rate, we  run the given configuration for $100\, P_{AB,0}$, where $ P_{AB,0}$ is the initial orbital period of the outer binary. The  orbital period of the outer binary may change slightly but the period stayed within 2\% of its initial value in our standard configuration. After each orbit we  record the change in $\varpi_{AB}=\omega_{AB}+\Omega_{AB}$ and use this to  calculate the average $\dot \varpi_{AB}$.  The numerically averaged precession rate is in good agreement with the analytical form as will be discussed later.} For these standard parameters, the precession is in a prograde direction for $i_{AB}<55^\circ$  and $i_{AB}>125^\circ$ and retrograde for $55^\circ<i_{AB}<125^\circ$.  We discuss this further in Section~\ref{subsec:eA}.

\subsection{Analytic model}
\label{sec:analytic1}

In the quadrupole approximation, with the approximation that the angular momentum of the outer binary is much larger than that of the inner, the precession rate for the longitude of the periapsis of the companion for the triple
is given by
\begin{multline}
\dot \varpi_{AB} =
\left(\frac{3 \sqrt{G} }{4} \right)
\left(\frac{m_{Aa} m_{Ab}(m_{AB})^{1/2}}{(m_A)^2}\right)\times\\
\left(\frac{a_A^2}{a_{AB}^{7/2}}\right)
\left(\frac{1}{(1-e_{AB}^2)^{2}}\right)F(e_A,i_{AB}),
\label{eq:efit}
\end{multline}
\citep[e.g.][]{Morais2012,Naoz2016,Lepp2023}
where 
\begin{multline}
    F(e_A,i_{AB})= (1+\alpha e_A^2)\left(\frac{3\cos(i_{AB})^2-1}{2}
    \right)\\
    +\frac{15}{4}e_A^2(1-\cos(i_{AB})^2)\cos(2\omega_A),
    \label{eq:rateF}
\end{multline}
 $\omega_A$ is the argument of the periapsis of the inner binary measured relative to the outer binary, and the parameter $\alpha=3/2$.  Note that equation (25) of \cite{Morais2012}
 contains a typo. That equation is for the longitude of the periapsis $\dot{\varpi}_2$, rather than the argument of periapsis $\dot{\omega}_2$.   Previously, we found that $\alpha=2$ works better for parameters near our standard model \citep[see][]{Lepp2023}. We further discuss this below in section~\ref{sec:scaling}.
This is consistent with \cite{Naoz2016} but following \cite{Morais2012} it assumes that the outer binary is a fixed plane.  This is a good approximation as long as the outer binary contains most of the angular momentum (i.e. when $a_A \ll a_{AB}$ and $m_B \gtrsim m_A$).

The second term in equation~(\ref{eq:rateF})  depends on $\omega_A$. This term is zero for low mutual binary inclinations $i_{AB}$ and for circular orbit inner binary, $e_A =0$. In these two cases we can ignore this term, but we discuss it further in Section~\ref{subsec:eA}.  The first term in equation~(\ref{eq:rateF}) goes to zero when $\cos(i_{AB})^2=1/3$, where $i_{AB}= 55^\circ$ and $ 125^\circ$.

In the $n$-body simulations (Fig.~\ref{fig:maxe}), we saw that the average mutual inclination and inner binary eccentricity are close to their initial values. Therefore in the analytic model  we use   the initial values.  While this approximation is good for inclination, it is less so for eccentricity. However, the analytical formula is not very sensitive to eccentricity at lower eccentricities and so we find this works well. The lower panel of Fig.~\ref{fig:maxe} shows the analytic curve from equation~(\ref{eq:efit}) that is in good agreement with the numerical simulations for  the average apsidal precession rate over the entire range of initial mutual inclinations. The apsidal precession rate over an individual orbit varies over a broad range as the inner binary undergoes KL oscillations but the precession rate averaged over a large number of orbits is consistent with our analytical approximation.  Both $e_A$ and $i_{AB}$ vary during these oscillations but their effects on the precession rate tend to mostly cancel each other when averaged over a complete cycle.  This works less well for larger initial values of $e_A$ and this is  why our analytical curve doesn't work as well for $e_A >0.6$.  When the inner binary is also eccentric, \SL{this analytical expression only works when the initial value of $\omega_A$ is an} integer multiple of $90^\circ$ as discussed below in section \ref{sec:eA}.

\subsection[Apsidal precession rate scaling]{Apsidal precession rate scaling with $e_A$}
\label{subsec:eA}

\begin{figure*}
\begin{tabular}{c c}
\textbf{Case A}&\textbf{Case B}\\
\includegraphics[width=.99\columnwidth]{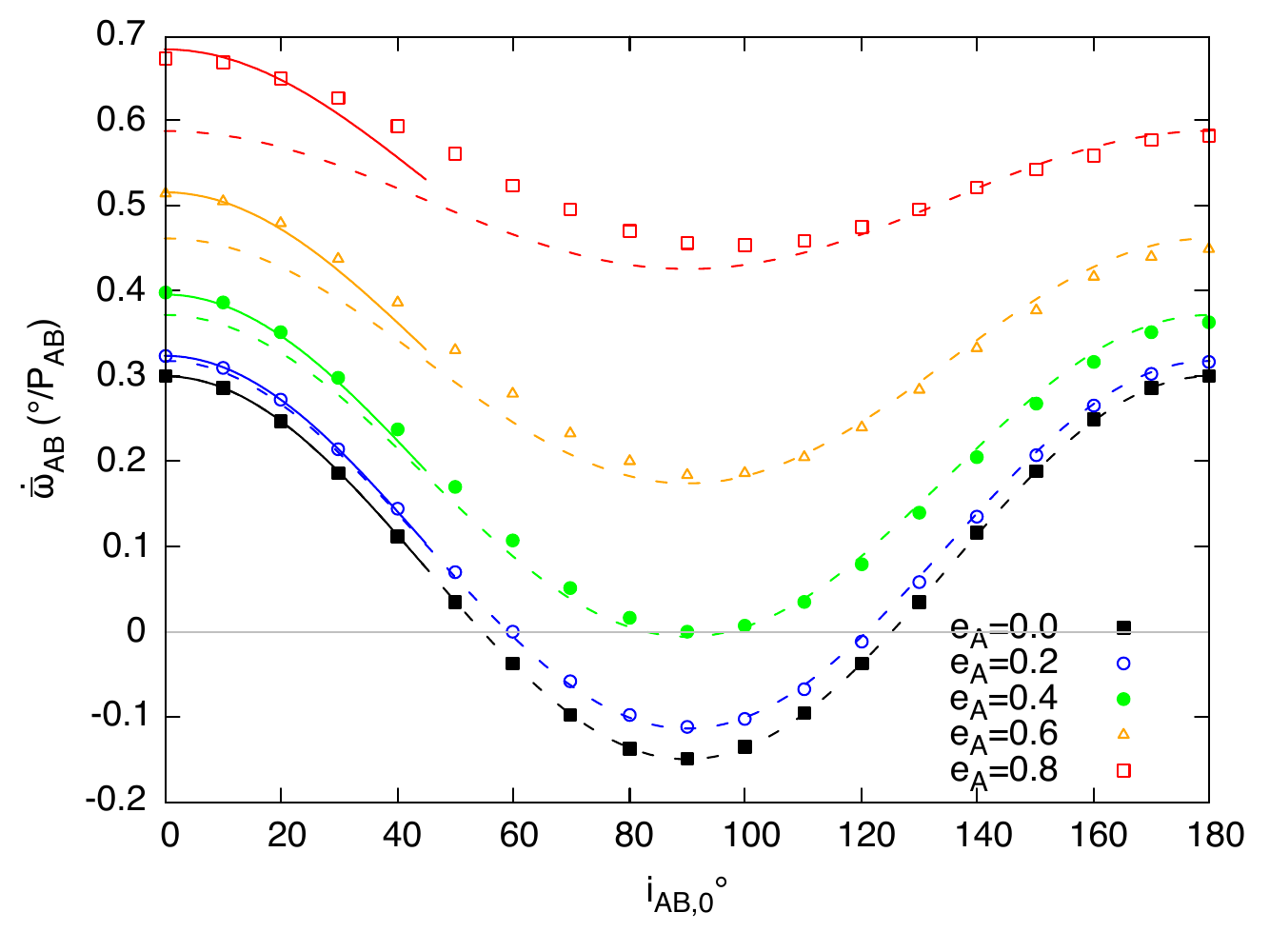}&
\includegraphics[width=.99\columnwidth]{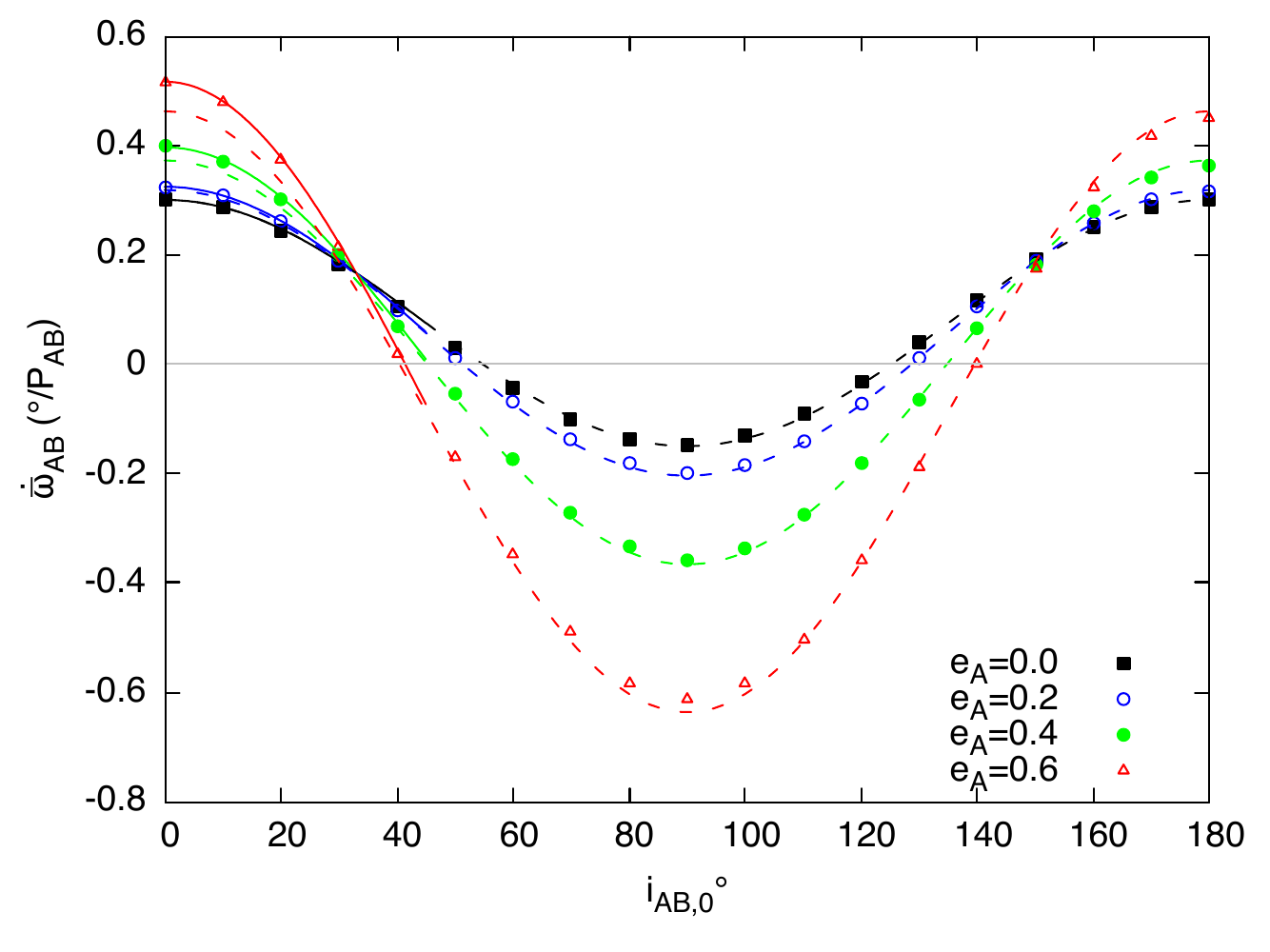}\\
\textbf{Case C}&\textbf{Case D}\\
\includegraphics[width=.99\columnwidth]{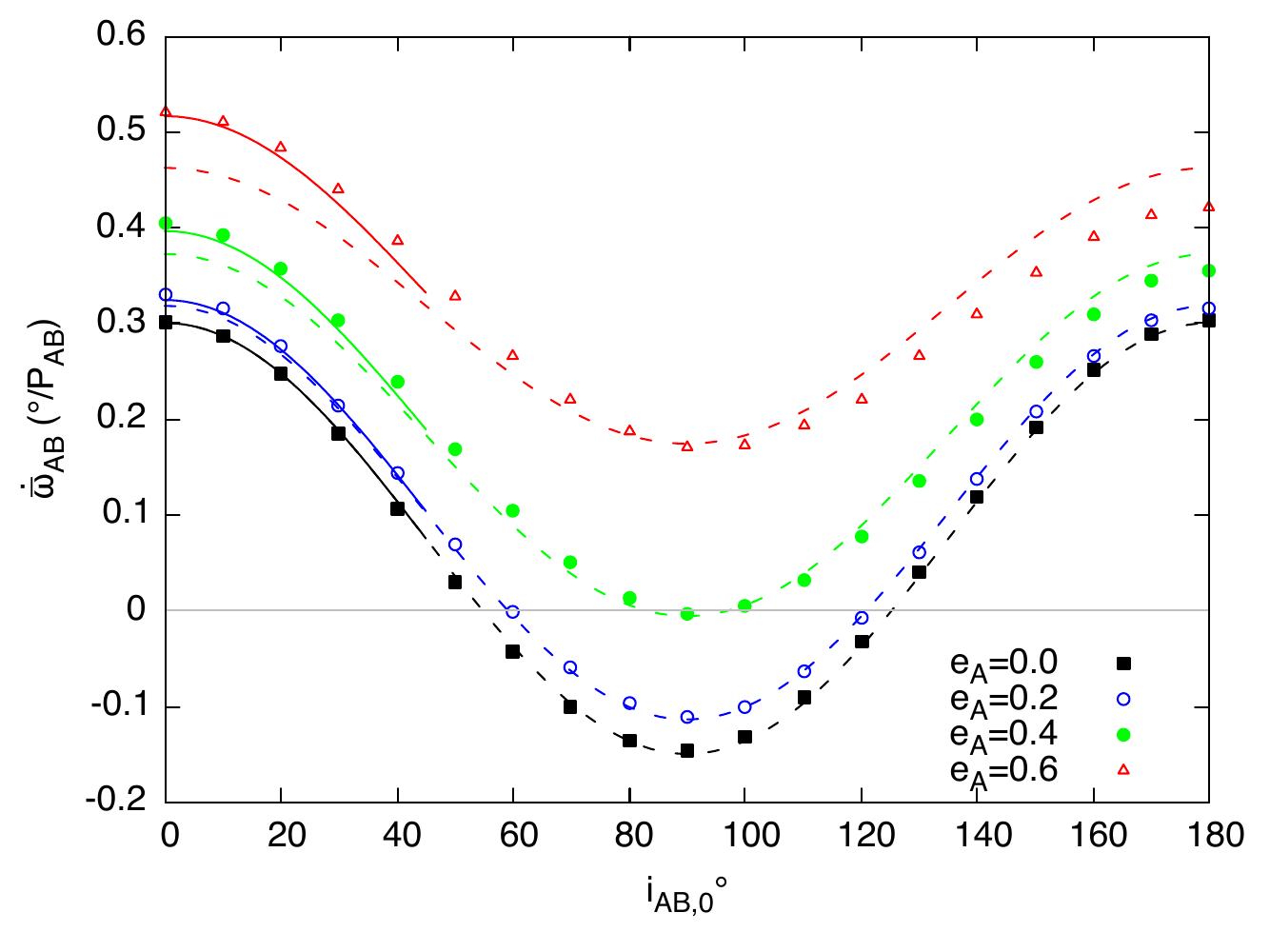}&
\includegraphics[width=.99\columnwidth]{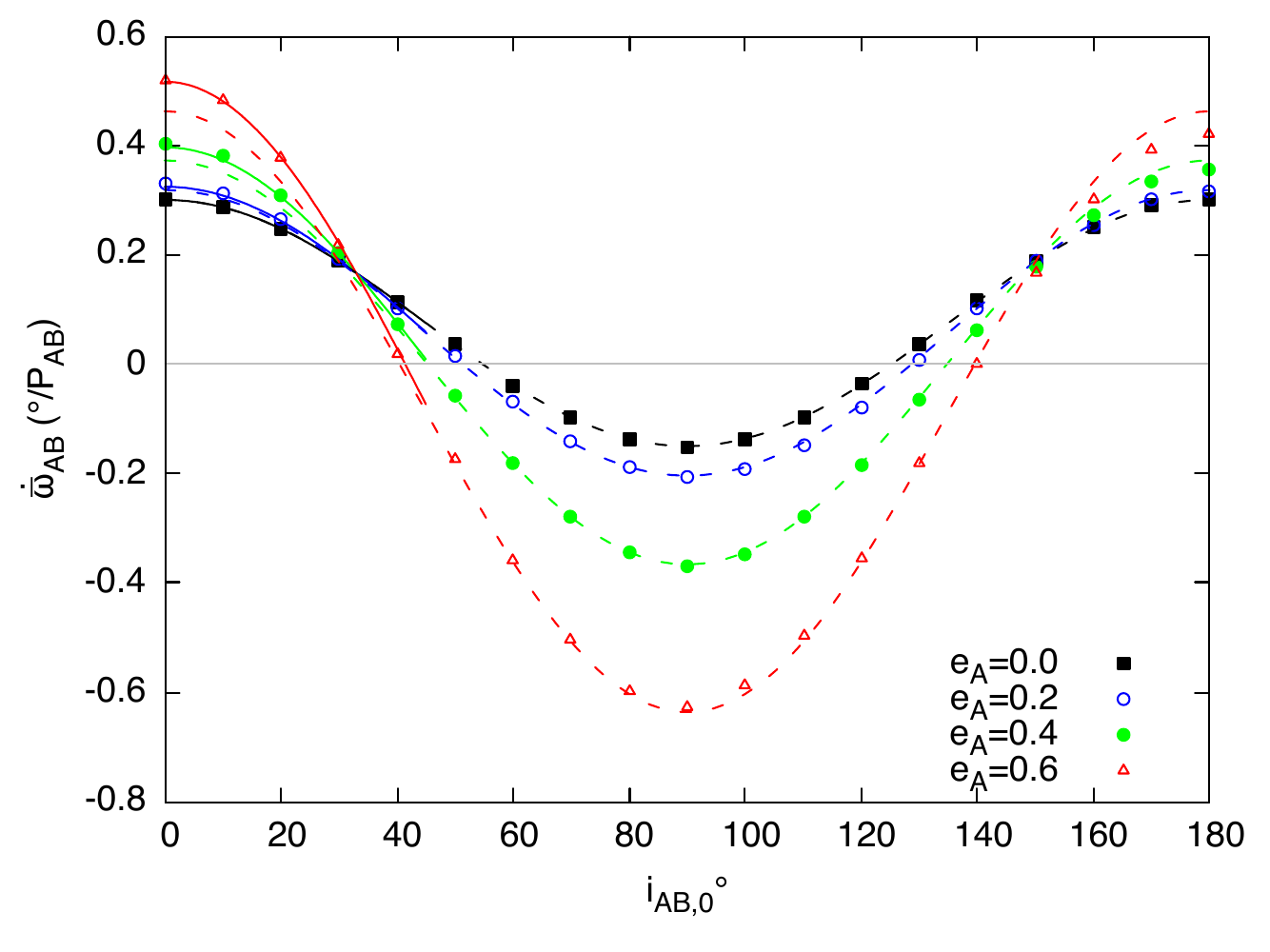}\\
\end{tabular}
\caption{The apsidal precession rate for our standard paramaters for the four cases in Table~\ref{tab:cases} with varying $e_A$. The precession rate is averaged over 100 initial periods of the outer binary and the units are degrees per $P_{AB,0}^{-1}$.  The analytic curve is shown as dashed lines for $\alpha=1.5$ and as solid lines for $\alpha=2$. The $\alpha=2$ curve is only shown for inclinations less than $40^\circ$ where it is applicable. The top left panel is case A, top right is case B, bottom left is case C and bottom right case D.}
\label{fig:prec2}
\end{figure*}

For the evaluation of $F(e_A,i_{AB})$ (equation~\ref{eq:rateF}) in \cite{Lepp2023}, there were two approximations made.  First we ignored the second term in this equation. This works because the second term goes to zero for both low mutual inclination of the binaries  and for small $e_A$.  In the first term, we modified the quadrupole approximation to have a parameter $\alpha$ which would be $\alpha=3/2$ but it was found $\alpha=2$ fit the numerical models better for our standard model. However, for larger separations $\alpha=3/2$ was a better fit.   

\SL{If $i_{AB}>40^\circ$ and $e_A>0.2$ there is good agreement between the  numerical results and the analytic apsidal precession rate with $\omega_A=0^\circ$ for cases A and C and $\omega_A=90^\circ$ for cases B and D. We  discuss this further in Section \ref{sec:eA}}. \SL{The analytic model assumes that $i_{AB}$ and $e_A$ remain fixed in time at their initial values.}
The analytical curve is shown in Fig.~\ref{fig:prec2} along with numerically calculated apsidal precession rates. The analytical function is shown as dashed lines and the points are numerical rates from {\sc rebound}. \SL{ The precession rate is calculated in the same way as in Figure~\ref{fig:maxe} except we average over a time of $1000\,P_{\rm AB,0}$ to ensure that several full KL cycles are included.} Cases~A and~C  both have $\omega_A=0^\circ$ initially and have very similar behavior. Cases~B and~D  both start with $\omega_A=90^\circ$ and are similar to each other. Near our standard configuration, the $\alpha$ parameter is better fit by $2$ than by $3/2$ and the solid lines show this fit.  Note that there is an asymmetry and so the coplanar prograde precession rate can be slightly faster than the retrograde coplanar apsidal precession rate for high eccentricity of the inner binary.  The fit is worse for large $e_A$ near retrograde for cases C and D.  
This may be due to higher order terms similar to our heuristic $\alpha=2$ for low inclination orbits. Finally, as can be seen in the first panel, the fit gets also gets worse as the eccentricity is increased to $e_A=0.8$ as discussed in the previous subsection.
For the rest of the paper we restrict ourselves to $e_A \le 0.6$.

\subsection{Critical mutual inclination for zero apsidal precession rate}

\SL{Solving equation~(\ref{eq:efit}) for where the triple star apsidal precession rate is zero, we} define the critical mutual inclination, 
\begin{equation}
\cos i_{\rm c}= \sqrt{\frac{2(1+\alpha e_A^2)-15 e_A^2 \cos(2\omega_A)}{6(1+\alpha e_A^2)-15 e_A^2 \cos(2\omega_A)}}.
\label{eq:ic}
\end{equation}
In the case of a zero eccentricity inner binary, this simplifies to $\cos i_{\rm c}=1/\sqrt{3}$ and so $i_{\rm c}=55^\circ$ and $i_{\rm c}=125^\circ$. For these two mutual inclinations, the triple star does not undergo apsidal precession and the dynamics of a circumtriple test particle would be the same as one orbiting around an equivalent binary star. In the next Section, we examine the dynamics of circumtriple particle orbits.

\begin{figure}
\includegraphics[width=.99\columnwidth]{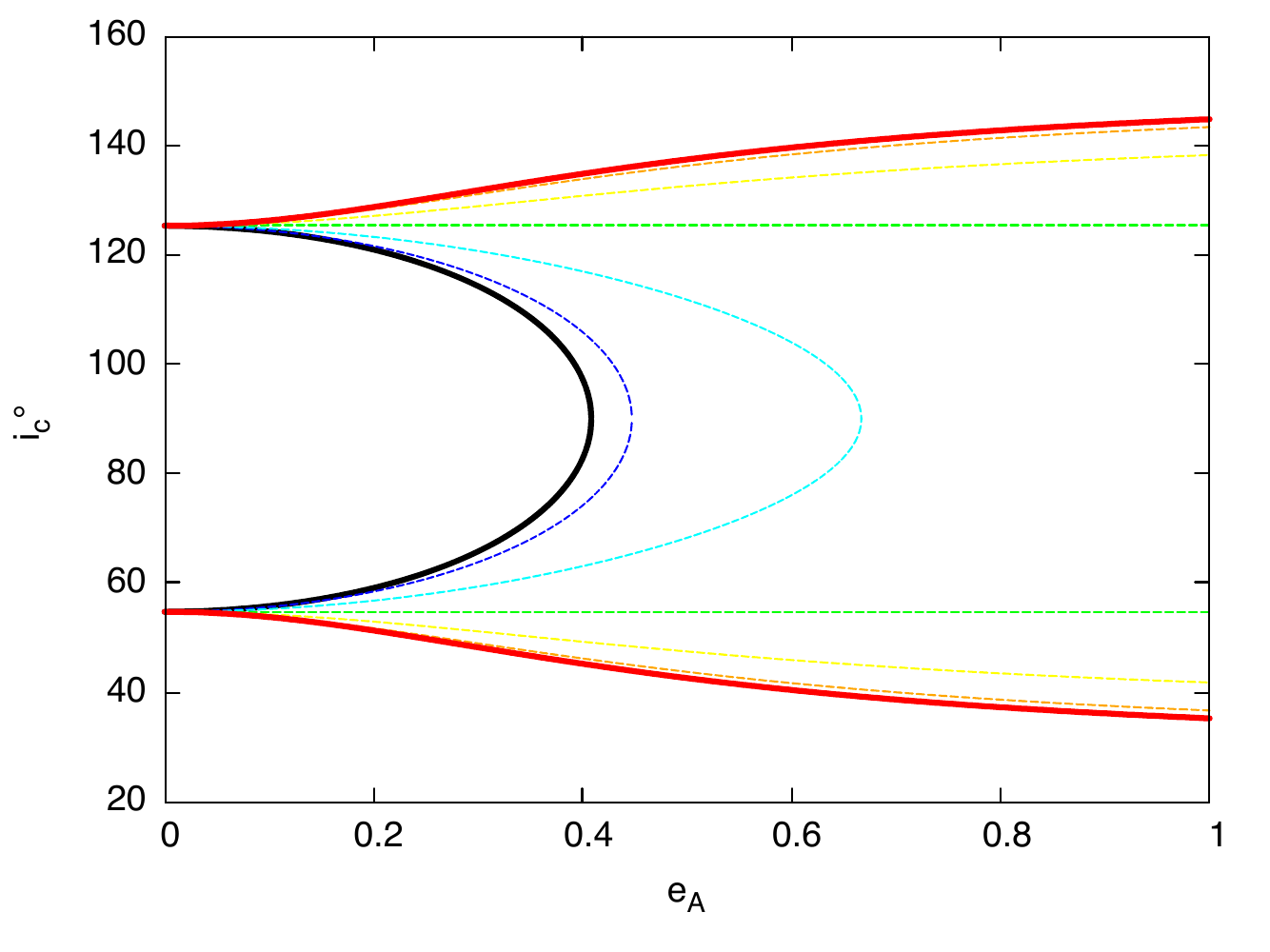}
\caption{The critical inclination, $i_{\rm c}$, where $\dot \varpi_{AB}=0$ as a function of $e_A$. The  curves for $\omega_A=0^\circ$ and $90^\circ$ in solid black and  red, respectively, represent \SL{the initial conditions for our simulations.}
The  curves for $\omega_A=15^\circ, 30^\circ, 45^\circ, 60^\circ$ and $75^\circ$ are dashed in blue cyan, green, yellow and orange, respectively. }
\label{fig:icrit}
\end{figure}
Figure~\ref{fig:icrit} shows the critical inclination as a function of $e_A$ from equation~(\ref{eq:ic}).  For the range where this equation is relevant we can use $\alpha=\frac{3}{2}$ and $\cos(\omega_A)=\pm 1$, where it is $+1$ when $\omega_A=0^\circ$ and $-1$ when $\omega_A=90^\circ$.  The black curve shows where $\omega_A=0^\circ$ and red for $\omega_A=90^\circ$ \SL{which represent the initial conditions for our simulations. The value of $\omega_A$ is not constant in time and is discussed in Section~\ref{sec:eA}.} 
At $e_A=0$ the critical inclinations are approximately $55^\circ$ and $125^\circ$ as mentioned above and for $e_A > 0.41$ there are no critical angles for the $\omega_A=0^\circ$ case.  Here the precession slows down but never passes through zero and changes direction.  For $\omega_A=90^\circ$ there are always two solutions which start at the same values as for $e_A=0$ but move slightly apart as the eccentricity of the inner binary increases.

\section{Circumtriple particle dynamics} 
\label{sec:orbits}

\begin{figure*}
\includegraphics[width=.66\columnwidth]{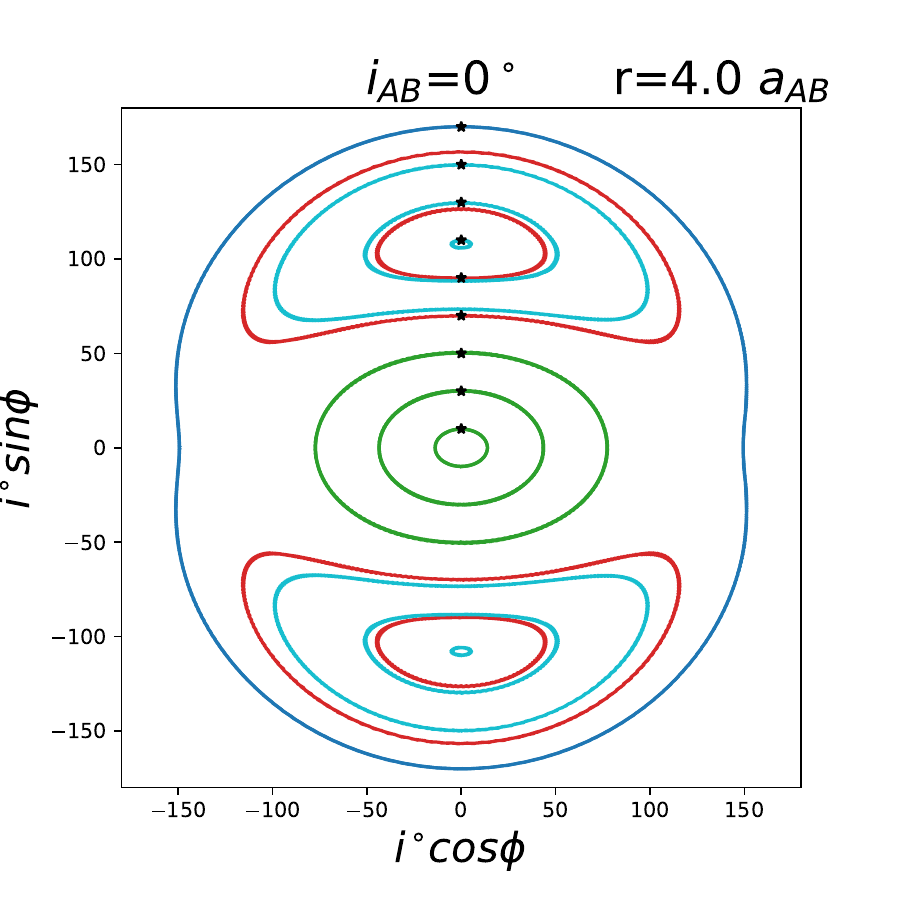}
\includegraphics[width=.66\columnwidth]{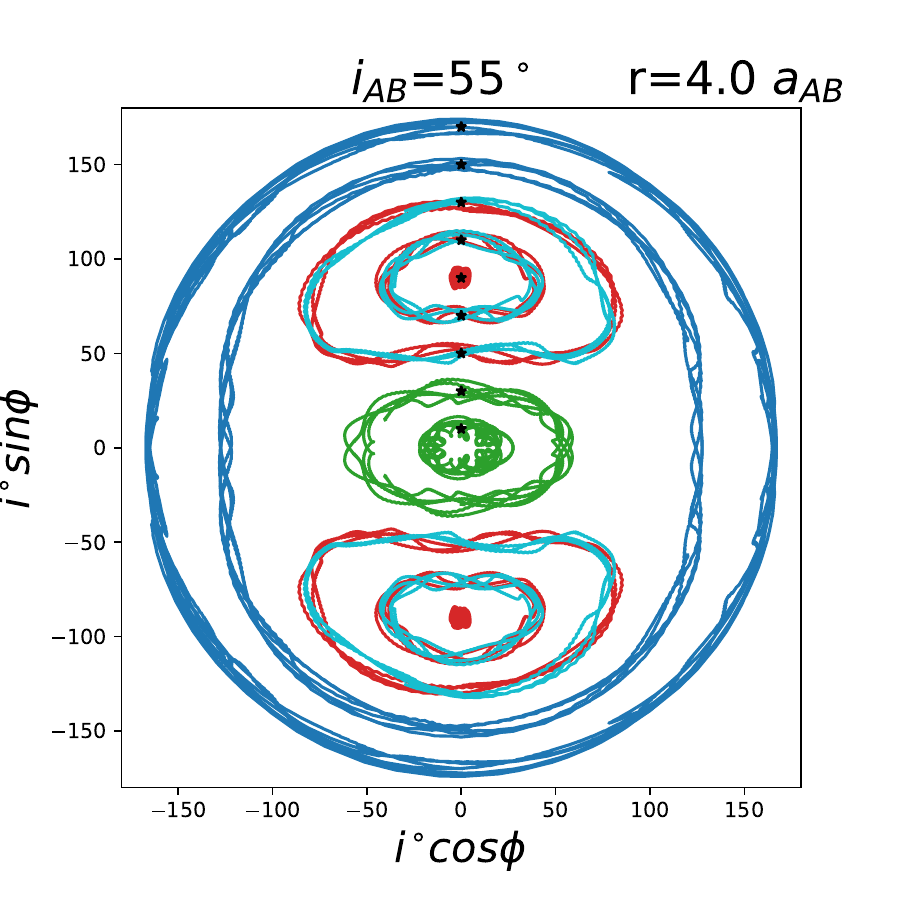}
\includegraphics[width=.66\columnwidth]{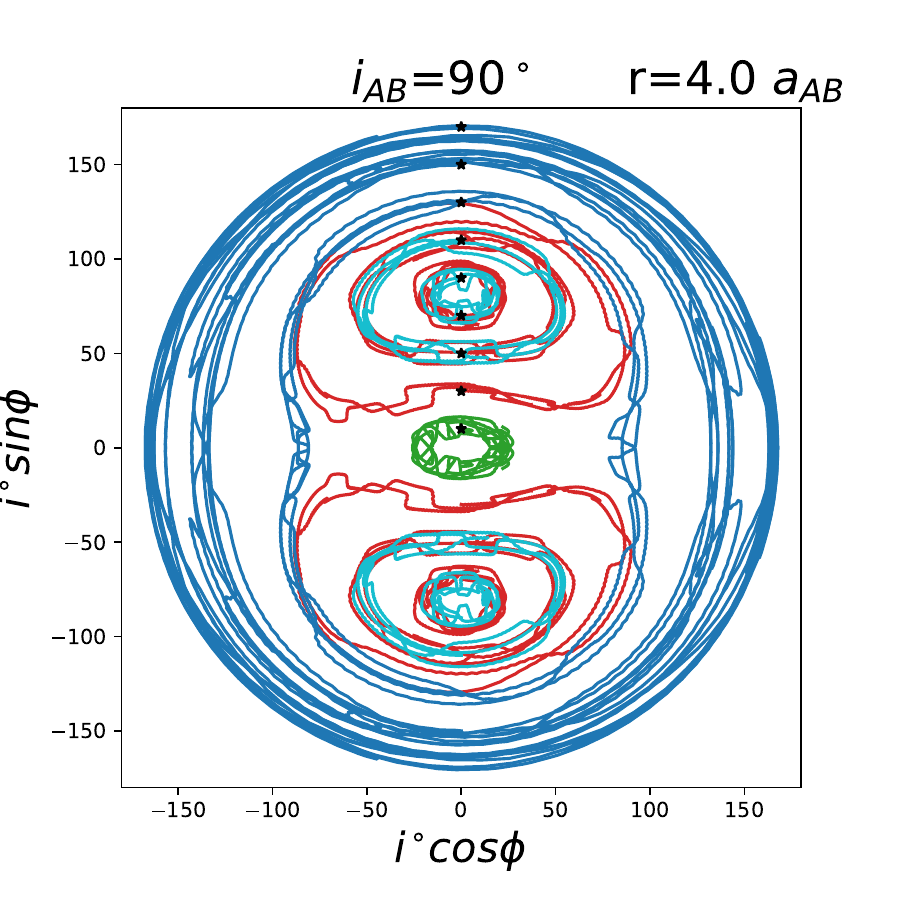}

\includegraphics[width=.66\columnwidth]{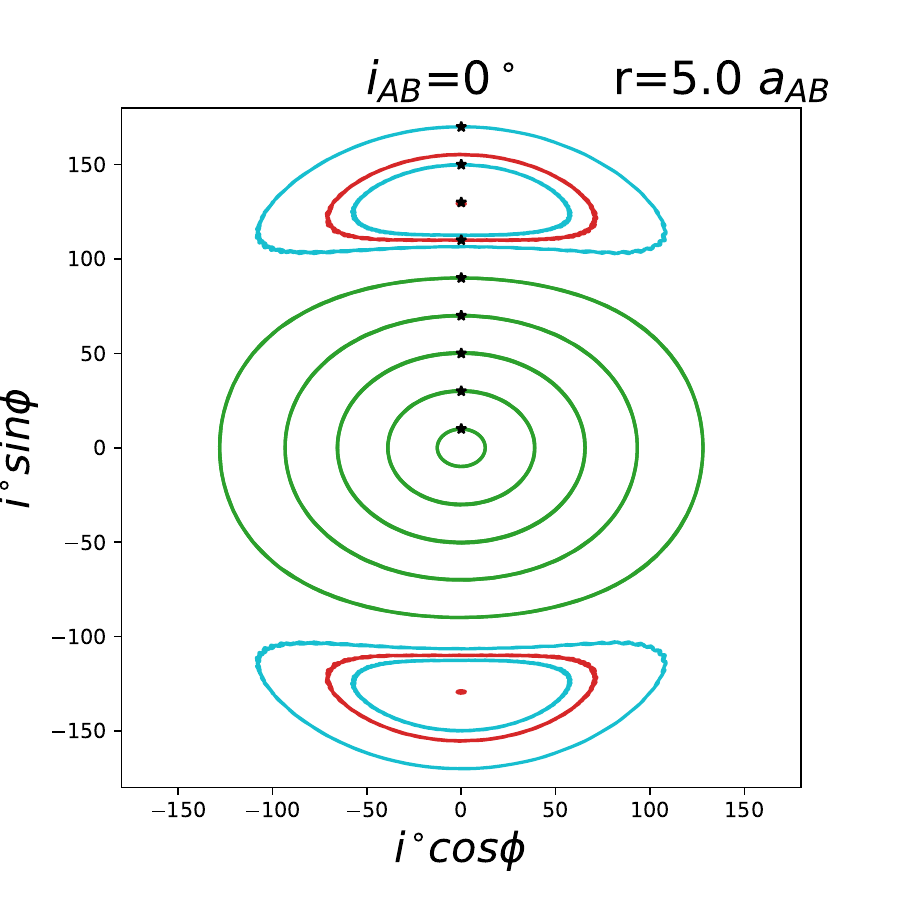}
\includegraphics[width=.66\columnwidth]{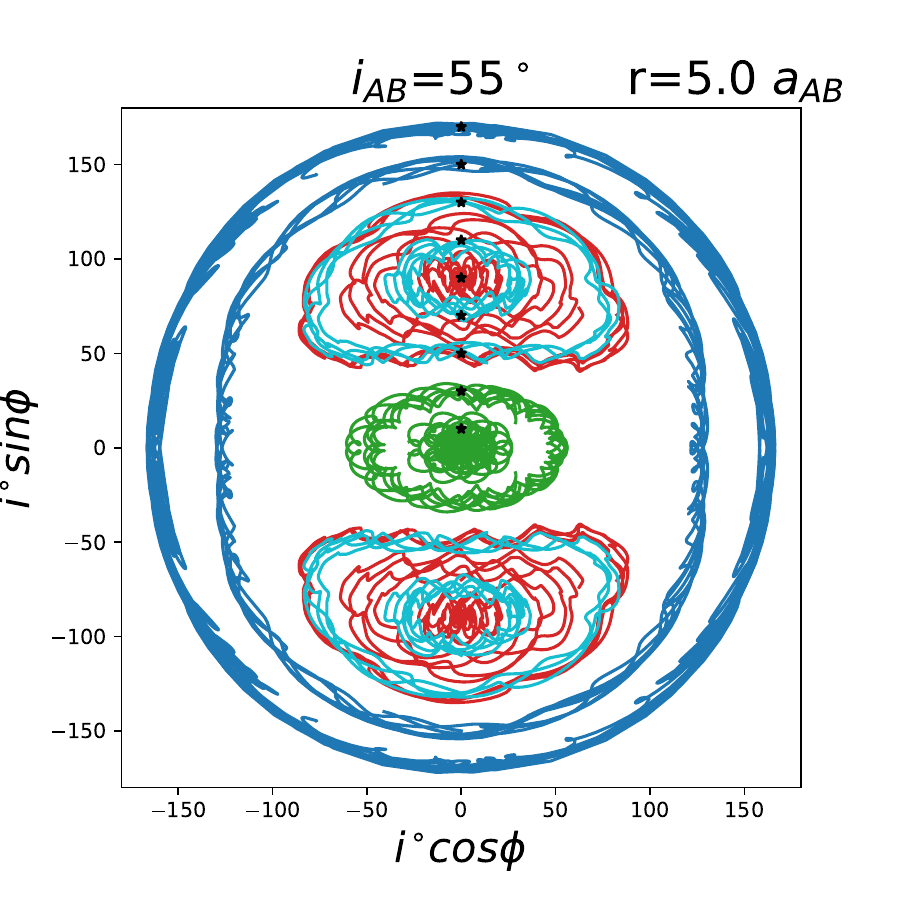}
\includegraphics[width=.66\columnwidth]{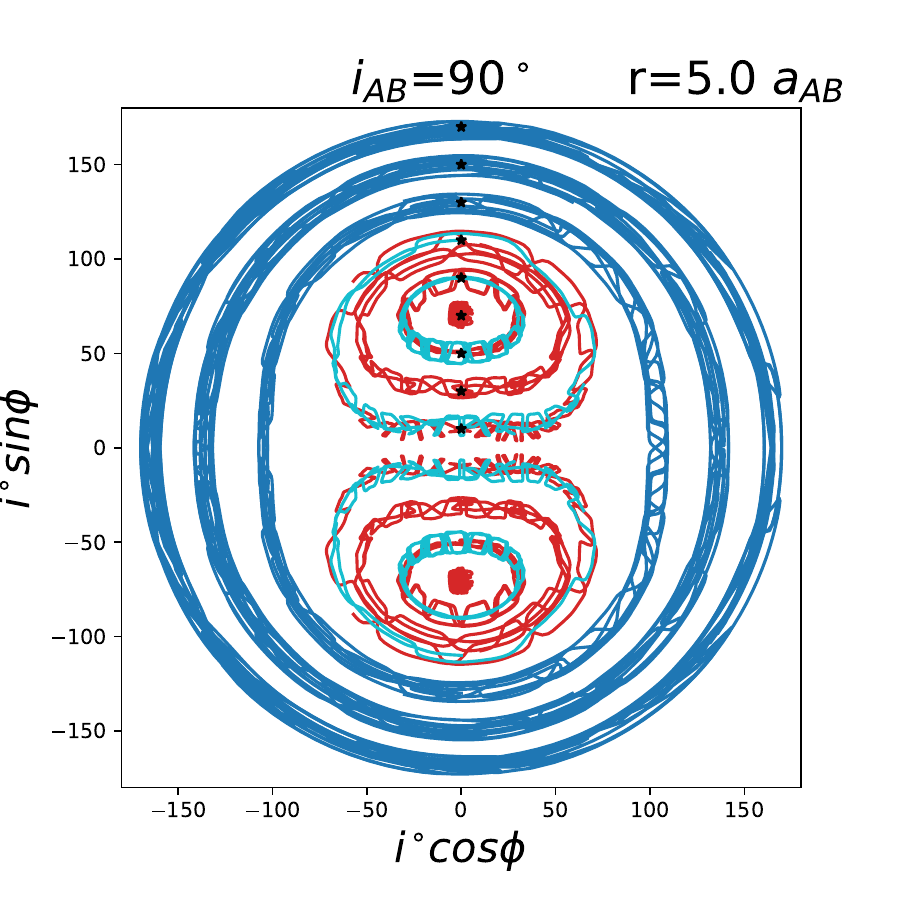}

\includegraphics[width=.66\columnwidth]{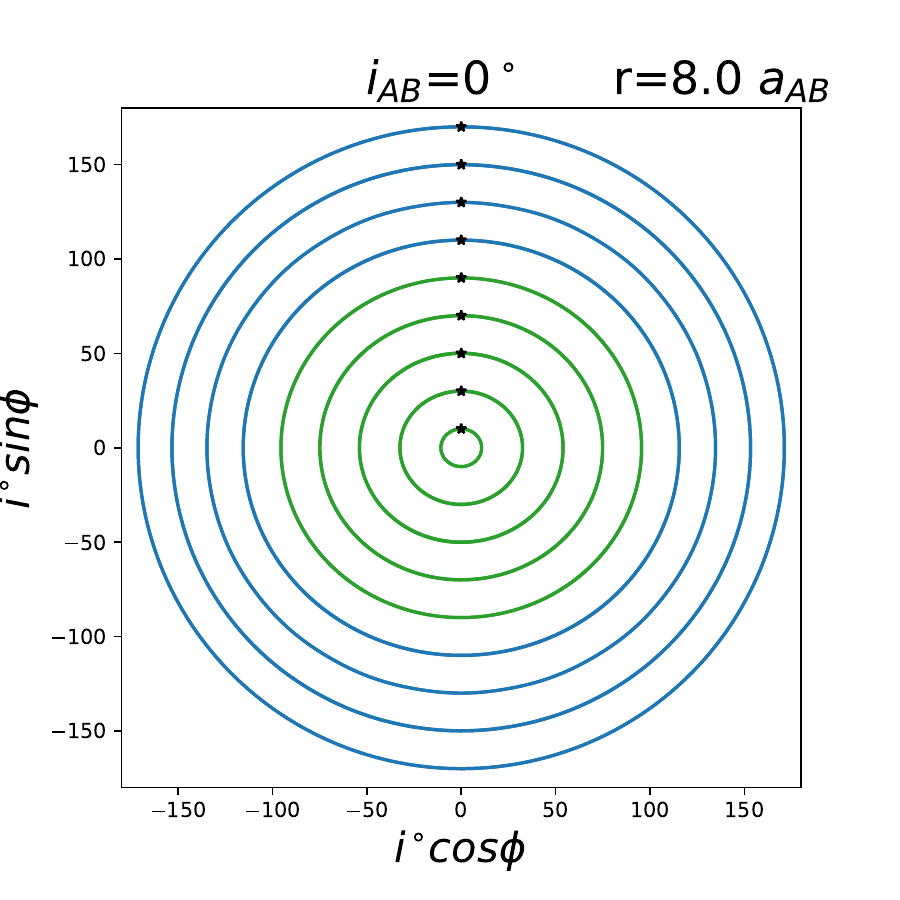}
\includegraphics[width=.66\columnwidth]{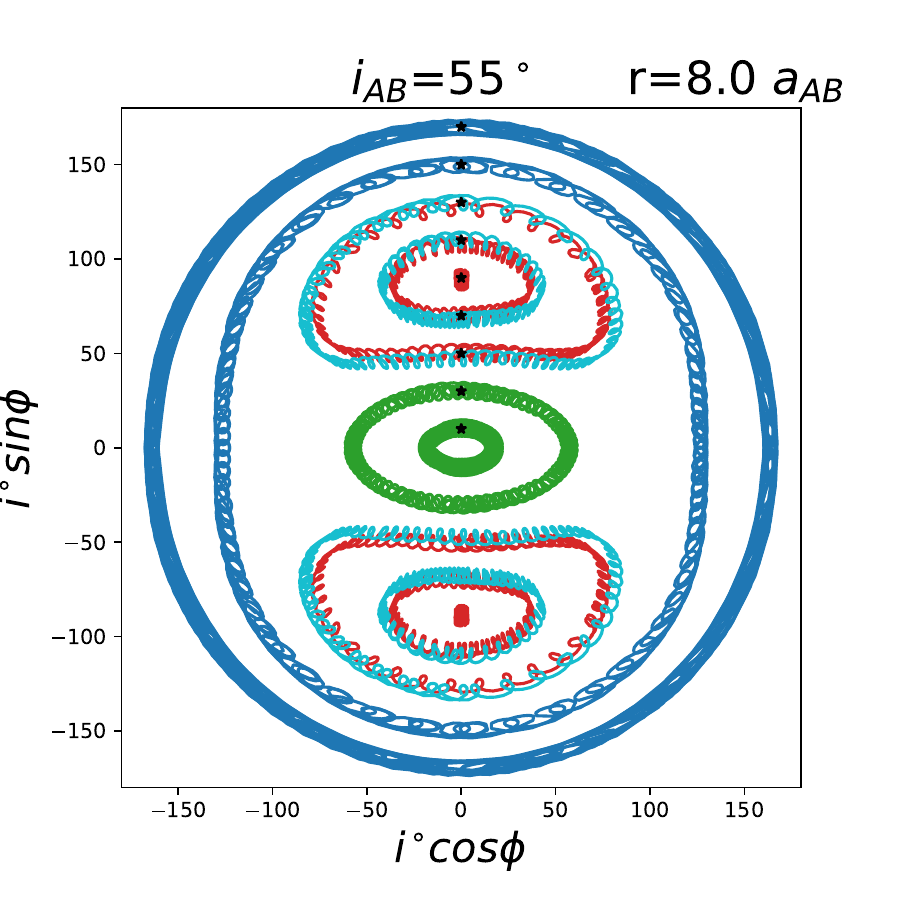}
\includegraphics[width=.66\columnwidth]{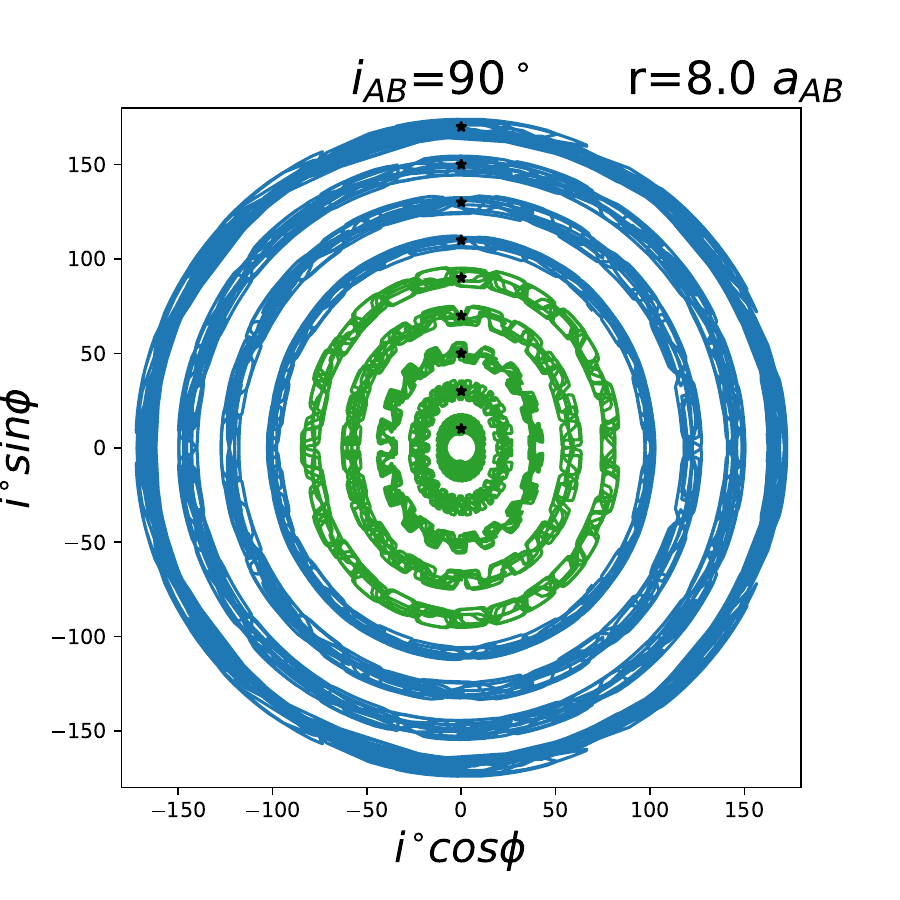}
\\
    \caption{\SL{The phase diagram  $(i\cos\phi,i\sin\phi)$ for circumtriple test particle orbits in the frame of the  outer binary for three mutual inclinations ($i_{AB}= 0^\circ, 55^\circ \text{ and } 90^\circ$) and  three test particle radii ($r= 4, 5 \text{ and } 8 a_{AB}$) for each inclination. The initial inclinations of the test particle relative to the outer binary are  $i=10, 30, 50, 70, 90, 110, 130, 150  \text{ and } 170^\circ$ with the starting point marked by a black star.  The orbits are colored green for prograde circulating orbits, red (cyan) for librating orbits that have an initial inclination lower (higher) than the stationary inclination, and blue for retrograde circulating orbits. Each test particle orbit is plotted twice,  normal and reflected about the x-axis, in order to show the librating region in the lower half. With increasing particle orbital radius, the stationary inclination increases for the coplanar ($i_{AB,0}=0^\circ$) case, decreases for the $i_{AB}=90^\circ$ case and  remains the same for $i_{AB}=55^\circ$. }}
    \label{fig:phase}
\end{figure*}

\begin{figure*}
\includegraphics[width=1.99\columnwidth]{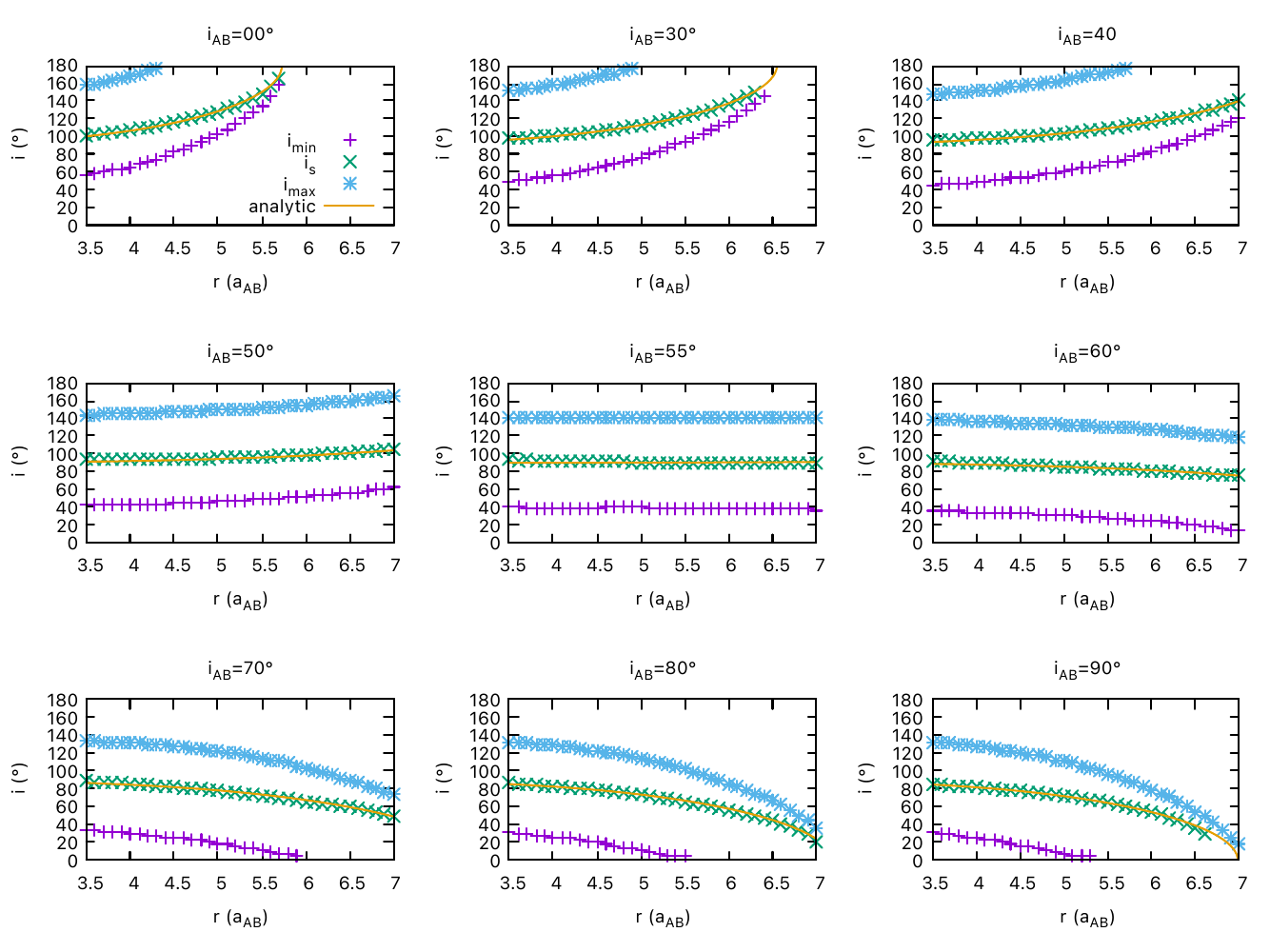}
\caption{The minimum initial inclination ($i_{\rm min}$, magenta) and maximum initial inclination  ($i_{\rm max}$, blue) for librating orbits and the stationary polar inclination (green) for varying inclinations or  $i_{AB}$ (in degrees) for our standard model.   The key in upper left panel applies to all nine panels. The $i_{AB}=10\textrm{ and }20^\circ$ cases are given in \cite{Lepp2023}. This configuration  is symmetric about $i_{AB}=90^\circ$ and so for initial inclinations between $90^\circ$ and $180^\circ$ are the same as the $i_{AB}=180^\circ - i_{AB}$ panel. Shown are curves calculated for Case A but since the precession rate is the same in all four cases at $e_A=0$ the other cases come out the same.  }
\label{fig:libreg}
\end{figure*}

We now consider test particle orbits around the triple star. The test particles begin in a circular orbit about the triple star at orbital radius, $r$ and at an inclination $i$ (with respect to the orbital plane of the outer binary). 
The inclination between a test particle orbit and the outer binary is given by
\begin{equation}
i = \cos^{-1}(\hat{\bm{l}}_{\rm AB}\cdot \hat{\bm{l}}_{\rm t})\,,
\end{equation}
where $\hat{\bm{l}}_{\rm AB}$ and $\hat{\bm{l}}_{\rm t}$ are the unit vectors in the direction of the angular momentum of the outer binary and the test particle respectively.
The nodal  phase angle of the test particle is 
defined as the angle measured relative to 
the eccentricity vector of the outer binary and is 
given by 
\begin{equation}
        \phi = \tan^{-1}\left(\frac{\hat{\bm{l}}_{\rm t}\cdot (\hat{\bm{l}}_{\rm AB}\times 
    \hat{\bm{e}}_{\rm AB})}{\hat{\bm{l}}_{\rm t}\cdot \hat{\bm{e}}_{\rm AB}}\right) + 90^\circ 
\end{equation}
\citep{Chen2019,Chen2020e}, where  $\hat{\bm{e}}_{\rm AB}$ is the  unit eccentricity vector of the outer binary. The test particle orbits begin with the initial angular momentum in the plane defined by the outer binary angular momentum  vector and its eccentricity vector.  Thus we have an initial inclination $i$ and the initial phase angle is $\phi=90^\circ$.

Depending on the initial inclination, the test particle orbit undergoes nodal precession that is either circulating or librating.  Circulating orbits are those in which the angular momentum vector of the test particle orbit precesses about the angular momentum vector of the outer binary's  angular momentum vector.  At higher inclinations the test particle orbits can librate, meaning that the angular momentum vector precesses about a stationary inclination. Around a binary star, the stationary inclination is $i=90^\circ$ and the librating nodal precession is centered on the eccentricity vector of the binary \citep{Verrier2009,Farago2010,Doolin2011,Naoz2017}. However, around a triple star, apsidal precession of the outer binary leads to the stationary inclination moving away from $90^\circ$ \citep{Lepp2022,Lepp2023}.

\subsection{Types of orbits}

\SL{
 Figure \ref{fig:phase} shows phase diagrams in the $(i\cos\phi,i\sin\phi)$ plane for some circumtriple test particle orbits.  The triple star has our standard parameters for three different initial mutual inclinations $i_{AB}=0^\circ$, $55^\circ$ and  $90^\circ$  with  test particle orbital radii of $r= 4$, $5$ and  $8\, a_{AB}$. The initial test particle inclinations relative to the outer binary are  $i=10^\circ$, $30^\circ$, $50^\circ$,  $70^\circ$, $90^\circ$, $110^\circ$, $130^\circ$, $150^\circ$ and  $170^\circ$. 
}
\SL{At all three mutual inclinations at the closest radius have a prograde circulating region (green), a librating region (red and cyan) and a retrograde circulating region (blue).  For larger test particle orbital radius, the librating region moves to higher inclination for  $i_{AB}=0^\circ$, lower inclination for $i_{AB}=90^\circ$ and stays the same for the $i_{AB}=55^\circ$.   }  
\SL{The triple stars in the $i_{AB}=55^\circ$ and $90^\circ$ cases undergo significant KL oscillations and the phase diagrams have the additional small loops not seen in the $i_{AB}=0^\circ$ case.  
Despite the large changes to the triple star orbit during a KL cycle, the test particle orbits are still clearly characterized as librating or circulating.}

Fig.~\ref{fig:libreg} shows the critical initial inclination of the test particle orbit for librating orbits (bounded by the blue and magenta lines) as well as stationary inclination (green lines) as a function of orbital radius of the test particle for the standard model parameters  that includes $e_A=0$.  \SL{ 
A test particle orbit is considered to be librating if the parameters return to their starting values without moving to $\sin\phi<0$.
The  initial inclination is stepped in increments of $1^\circ$ and we find the lowest and highest inclinations that librate.  The stationary state is found where the librating orbit stays at nearly the same inclination.   The time that the simulation is run is chosen to be at least one libration time scale.}
\SL{The libration time for a test particle is significantly longer than the KL oscillation time for the three star system.  The $r=4 \, a_{AB}$ systems were run for $2000\,P_{AB,0}$  which about 13 KL cycles for the $i_{AB}=55^\circ$ case and 22 KL cycles for the $i_{AB}=90^\circ$ case. The  $r=5 a_{AB}$ simulations were run for $5000\,P_{AB,0}$  and the  $r=8\, a_{AB}$ simulations were run for $10,000\,P_{AB,0}$. }

In the absence of any apsidal precession, there is a prograde circulating region for approximately initial inclinations of  $i< 40^\circ $, a librating region for  inclinations of  $40^\circ<i<140^\circ$ with  a stationary inclination of  $90^\circ$ and  a retrograde circulating region for  $i>140^\circ$.  At the smallest test particle radius shown ($r=3.5 \,a_{AB}$), all of the triple star mutual inclinations are similar to this.  
Fig.~\ref{fig:libreg} was calculated with case A but the precession rate is the same at $e_A=0$ for all four cases and so the other three cases produce similar figures.

For mutual inclination either close to prograde, $i_{AB}<55^\circ$, or close to retrograde, $i_{AB}>125^\circ$ the triple star undergoes prograde apsidal precession (see the lower panel of Fig.~\ref{fig:maxe}). The initial particle inclinations that librate moves to larger inclinations for larger particle orbital radius. At a critical radius, $r_{\rm c}$, the stationary inclination is $i_{\rm s}=180^\circ$, at which point there are no librating orbits for larger particle orbital radius.  This is the same as the behavior described in \cite{Lepp2023} for the coplanar triple star case.  

For $55^\circ<i_{AB}<125^\circ$,  the apsidal precession is retrograde and the stationary inclination and librating region decrease with particle orbital radius. The critical radius $r_{\rm c}$, outside of where there are no librating orbits, occurs where the stationary inclination is $i_{\rm s}=0^\circ$. For mutual inclination of approximately $i_{AB}=55^\circ$ and $125^\circ$, the apsidal precession rate goes to zero (see the lower panel of Fig.~\ref{fig:maxe}). For these mutual inclinations, the libration region is independent of the particle orbital radius in the quadrupole approximation that we apply. This is the same as the binary star case. This can be seen in the middle  panel of Fig.~\ref{fig:libreg}.   

The location of the stationary inclination and librating region depend on the precession rate which can vary as the inner binary undergoes a KL cycle.  This is particularly a problem for inclinations near $180^\circ$ or $0^\circ$ where the oscillation may mean that such a state doesn't exist for part of the cycle.  As such, in our automated search we have limited the search to tracking the stationary inclination for values between $10^\circ$ and $170^\circ$ and then extrapolated to $180^\circ$ and $0^\circ$ to get the numerically calculated values of $r_c$.

We note that with  our standard parameters, specifically, with $e_A=0$, the behaviour is the same for mutual inclination $i_{AB}$ and $180^\circ-i_{AB}$. However, we consider the case where $e_A$ varies in Section \ref{sec:eA}.

\begin{figure}
\includegraphics[width=0.99\columnwidth]{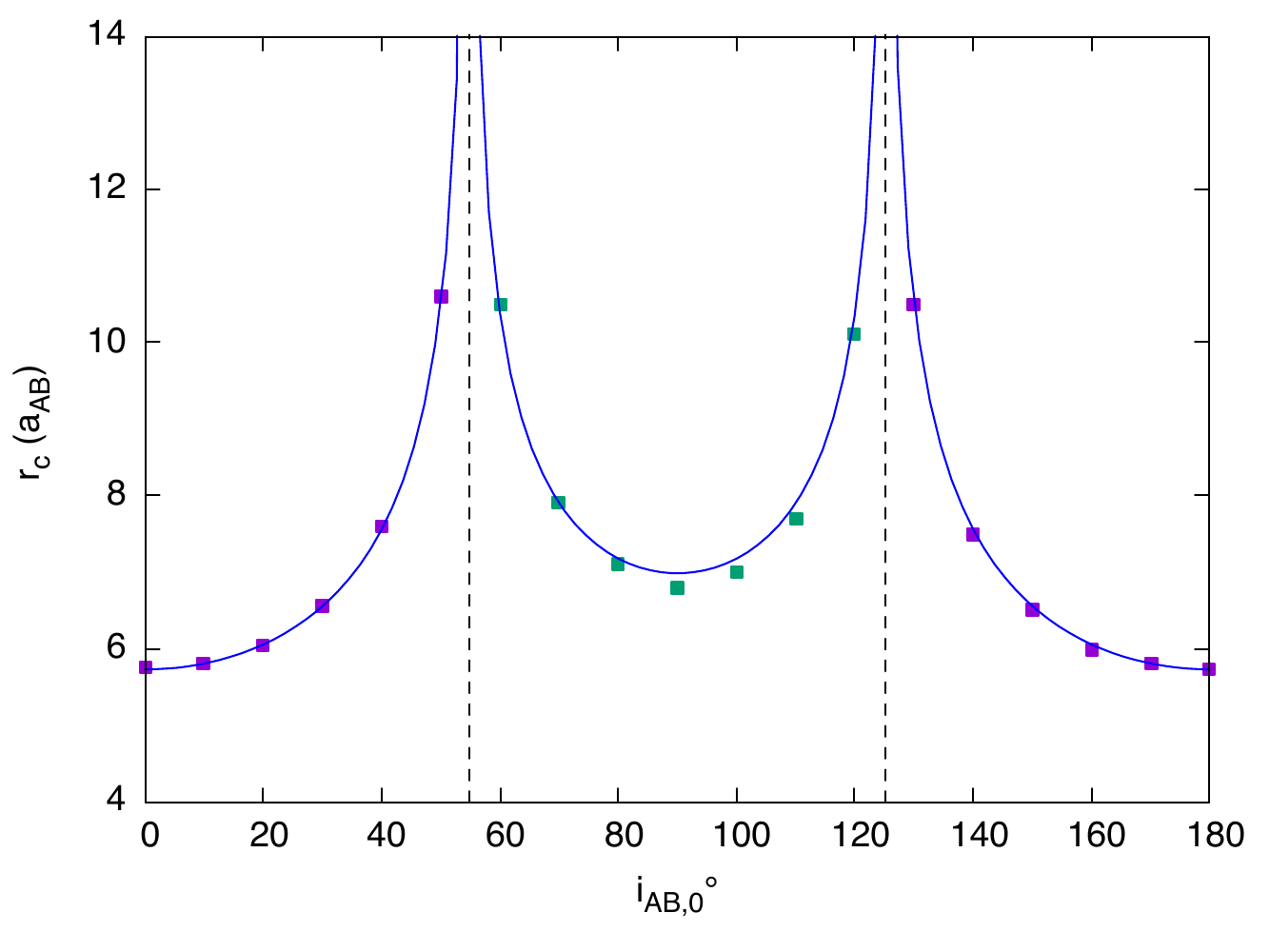}
\caption{The maximum radius for a librating orbit as defined by where the stationary inclination reaches either $180^\circ$ or $0^\circ$ \SL{for our standard triple star parameters}.  The analytic curves are from Section~\ref{sec:analytic}, $r_{\rm c}$ is calculated from the data in Fig.~\ref{fig:libreg} extrapolating where the stationary inclination is $i_{\rm s}=180^\circ$ or $0^\circ$. There are vertical lines at the critical inclinations found in equation~(\ref{eq:ic}) of  $i_{\rm c}=54.7^\circ$ and $125.3^\circ$. }
\label{fig:maxr}
\end{figure}

The critical radius outside of which there are no librating orbits is shown in Fig.~\ref{fig:maxr} as a function of the initial mutual inclination of the inner and outer binary.   
Note that the maximum radius goes to infinity at approximately $i_{AB}=55^\circ$ and $125^\circ$ as at these angles the apsidal precession rate goes to zero.  
This plot shows that librating orbits occur at larger particle orbital radius around hierarchical triple stars with mutual misalignments. We discuss the implications of this result in Section~\ref{conc}.


\subsection{Analytic Approximation to Circumtriple Orbits} 
\label{sec:analytic}

With the triple star precession rate, $\dot \varpi_{AB} $ given in equation~(\ref{eq:efit}), we can determine the circumtriple particle stationary inclination, $i_s$.  The stationary inclination occurs when the nodal precession rate of the test particle at a nodal phase of $\phi=90^\circ$ is equal  to the rate of change of the longitude of the periapsis of the binary \citep{Zanardi2018,Lepp2022,Lepp2023}.  This gives
\begin{multline}
    i_s=\cos^{-1}\left(
   -  \frac{m_{Aa}m_{Ab}m_{AB}^2}{m_A^3m_B}
     \left(\frac{r}{a_{AB}}\right)^{7/2}
    \times \right. \\
    \left.
    \left(\frac{a_A}{a_{AB}}\right)^2
   \frac{1}{(1+4e_{AB}^2)(1-e_{AB}^2)^2}\, F(e_A,i_{AB})
    \right).
    \label{eq:is2}
\end{multline}
A test particle at this inclination is stationary in the frame of the outer binary.  The analytic approximation to the stationary inclination is also shown in Fig.~\ref{fig:libreg}.  For  prograde apsidal precession of the outer binary, the stationary inclination increases with increasing test particle orbital radius and for a retrograde apsidal precession it decreases.  Librating orbits nodally precess around the stationary inclination and, as such, when the stationary inclination reaches  either $180^\circ$ or $0^\circ$ then the libration region can no longer occur.  This is the critical radius and it can be found by solving for $r$ when $i_s$ in equation~(\ref{eq:is2}) is either $180^\circ$ or $0^\circ$ depending on whether the apsidal precession is prograde or retrograde.  

The critical radius is given by 
\begin{multline}
    \frac{r_c}{a_{AB}}=\left(
    \frac{m_A^3m_B}{m_{Aa}m_{Ab}m_{AB}^2}
     \left(\frac{a_{AB}}{a_A}\right)^{2}   \times \right. \\ \left.
    \frac{(1-e_{AB}^2)^2(1+4e_{AB}^2)}{|F(e_A,i)|}
    \right)^{2/7}.
    \label{eq:rc}
\end{multline}
Note that $F$ also changes sign as the precession rate changes sign and so the absolute value around $F$ is required to ensure that the critical radius is always real.  This critical radius is plotted in Fig.~\ref{fig:maxr} and  provides a good fit over the entire range of initial binary mutual inclinations.  

\begin{figure*}
\includegraphics[width=1.1\columnwidth]{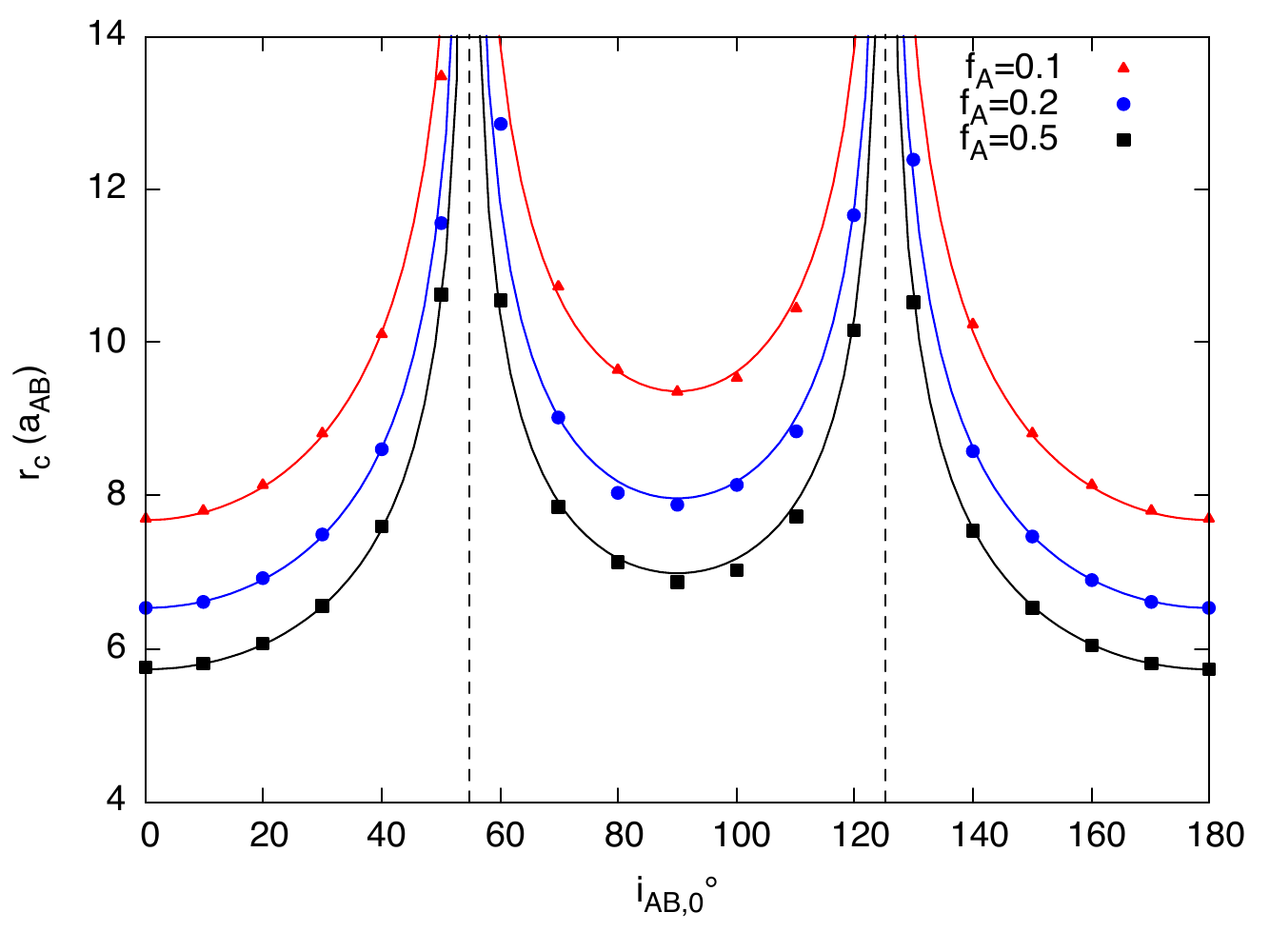}
\hspace{-0.25in}
\includegraphics[width=1.1\columnwidth]{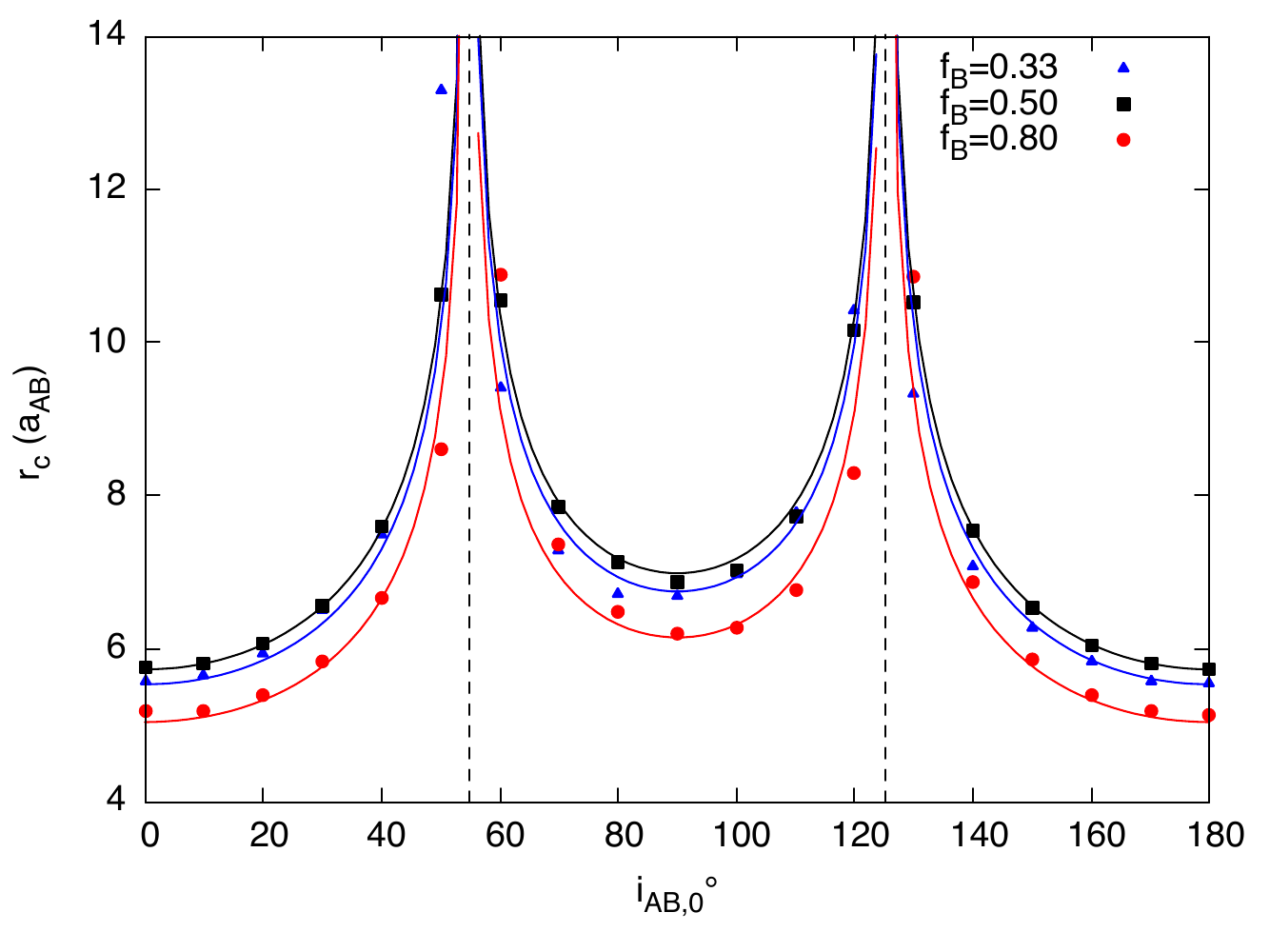}
\includegraphics[width=1.1\columnwidth]{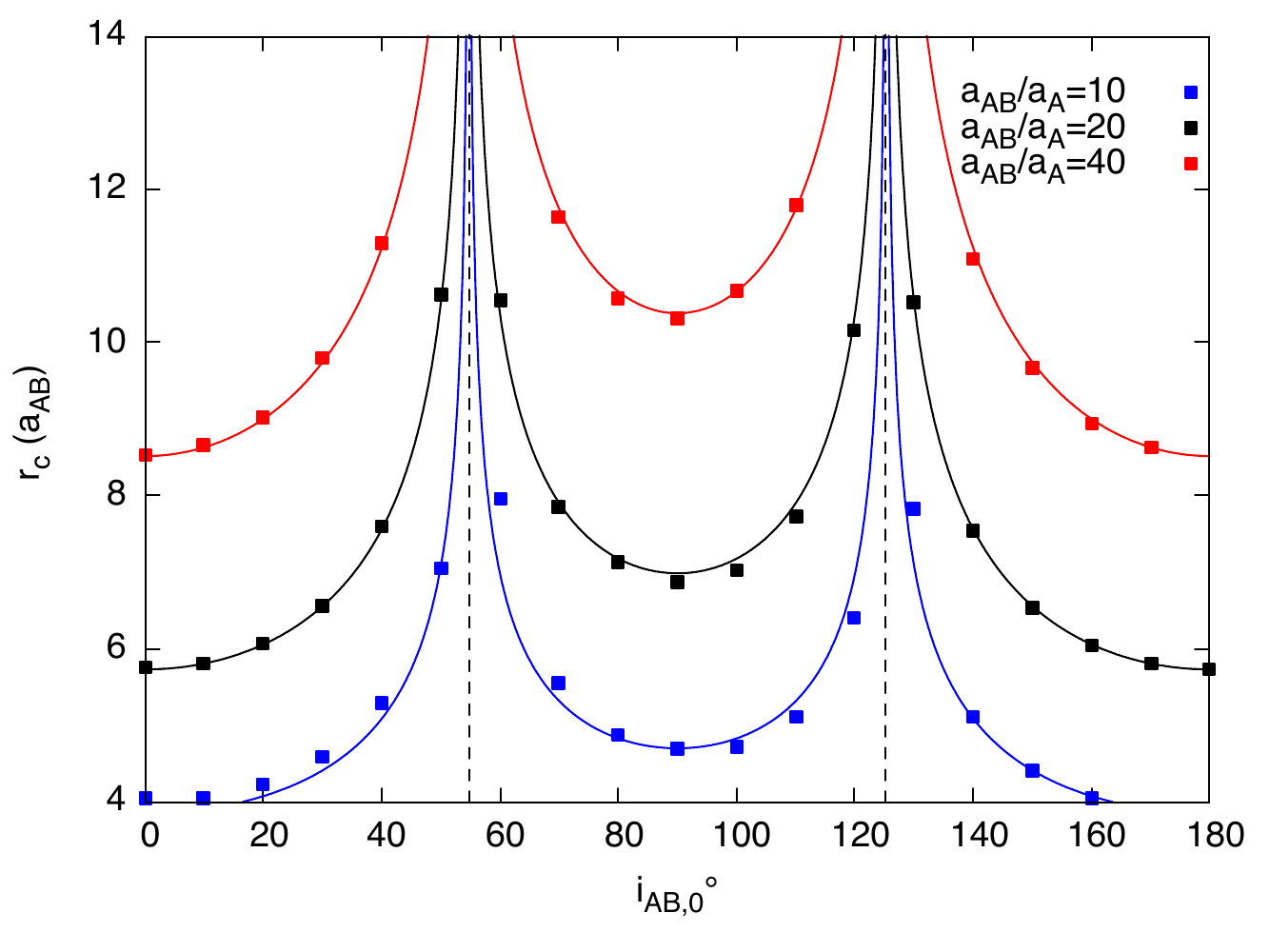}
\hspace{-0.25in}
\includegraphics[width=1.1\columnwidth]{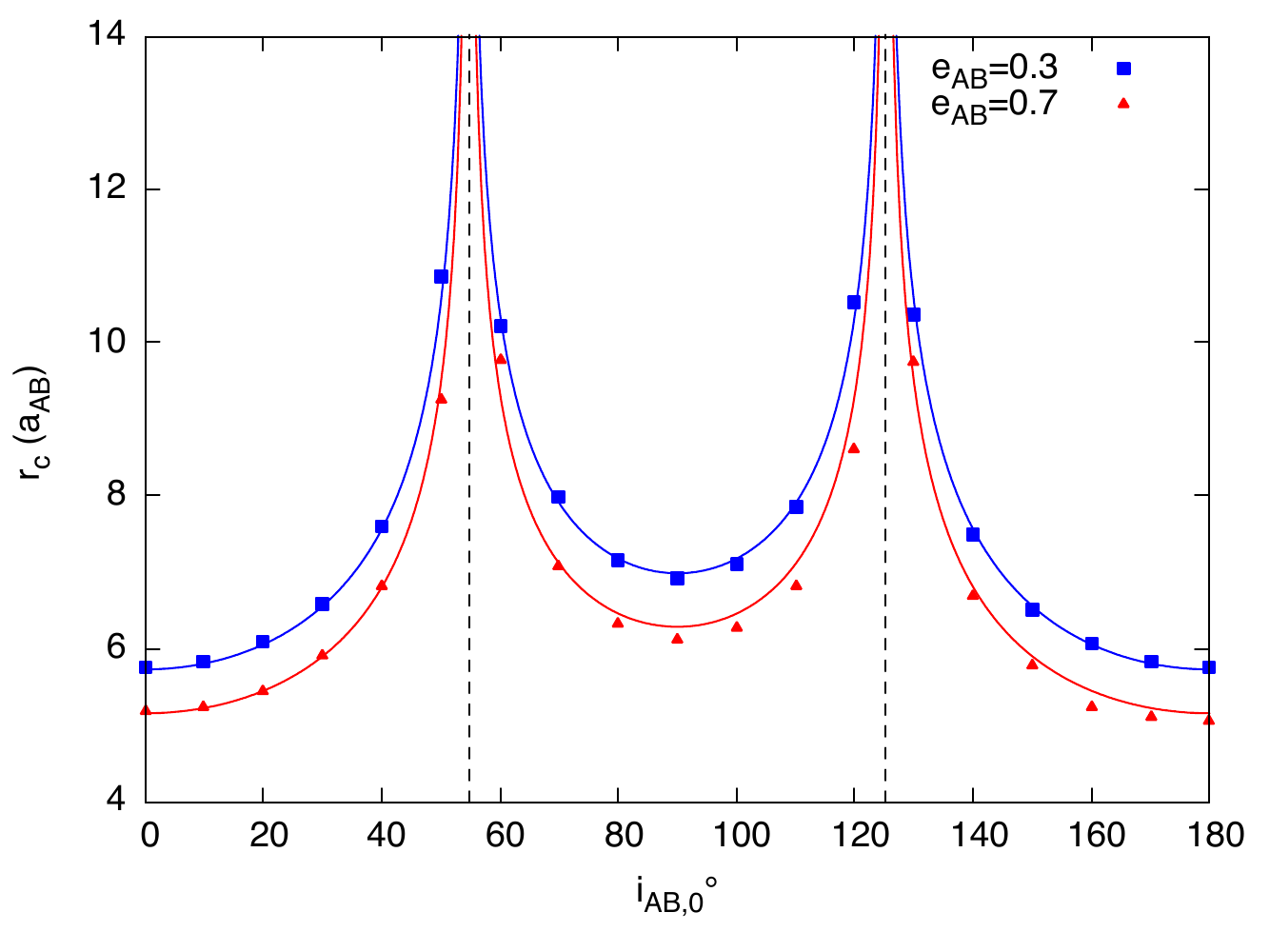}
    \caption{The critical radius $r_c$ vs the initial mutual inclination for a number of parameter changes.  The changes shown here $f_A$ (top left), $f_B$ (top right), $a_{AB}/a_A$ (bottom left) and $e_{AB}$ (bottom right). All of these just change an overall scaling factor but the functional form of the inclination dependence stays the same (since $e_A=0$ in all cases).  The blue lines show our analytical model. }
    \label{fig:scale1}
\end{figure*}

\subsection{Scaling with triple star parameters}
\label{sec:scaling}

The critical radius given in  equation~(\ref{eq:rc}) is identical to that in \cite{Lepp2023}, except for the absolute value sign around $F(e_A,i)$.  Therefore the same scalings that were used in \cite{Lepp2023} also apply here. For our standard parameters, $r_c/a_{AB}=5.73$, in agreement with Fig.~\ref{fig:maxr}. We can
then rewrite this equation as
\begin{equation}
    \frac{r_c}{a_{AB}}=5.73\, \frac{M \,A\, E}{F^{2/7}},
\end{equation}
where $M$, $A$ and $E$ are scaling functions for 
the radius in terms of mass, semi-major axis, and eccentricity of
the companion which have been normalized to one for our standard parameters. $F(e_A,i)$ is our same function which works for scaling since $F(0,0)=1$. 

As in \cite{Lepp2023} the scaling with masses is
\begin{align}
    M(m_{Aa},m_{Ab},m_B) & = \left( \frac{m_A^3m_B}{m_{Aa}m_{Ab}m_{AB}^2} \right)^{2/7}\\
    & = \left(\frac{(1-f_B)f_B}{(1-f_A)f_A}\right)^{2/7},
    \label{eq:mass}
\end{align}
 with semi-major axis is
\begin{equation}
    A(a_A,a_{AB}) =  \left(\frac{a_{AB}}{20\, a_A}\right)^{4/7},
\end{equation}
with the companion eccentricity is
\begin{equation}
        E(e_{AB}) =
        \left( \frac{8 (1-e_{AB}^2)^2(1+4e_{AB}^2)}{9} \right)^{2/7}.
\end{equation}
All of these scaling factors are unchanged from the coplanar triple star case.
However, the scaling  $F(e_{A},i)$  does require some modification.  The scaling used in \cite{Lepp2023} is still fine for low inclinations and works if $e_A$ is zero or small, but for higher inclinations and high  $e_A$ it requires some modification and is discussed below.

Fig.~\ref{fig:scale1} shows $r_c$ vs initial $i_{AB}$ for various values of the mass ratios, $f_A$ and $f_B$, the distance ratio $a_{AB}/a_A$ and the eccentricity of the outer binary $e_{AB}$.  In each of these cases our analytical model just introduces a scaling factor to $r_c$ but does not change the dependence on initial inclination.  The points represent $n$-body simulations for these parameters and they agree well with our analytical model.

\subsection[Scaling]{Scaling with $e_A$ }
\label{sec:eA}

\begin{figure*}
\begin{tabular}{c c}
\textbf{Case A}&\textbf{Case B}\\
\includegraphics[width=.99\columnwidth]{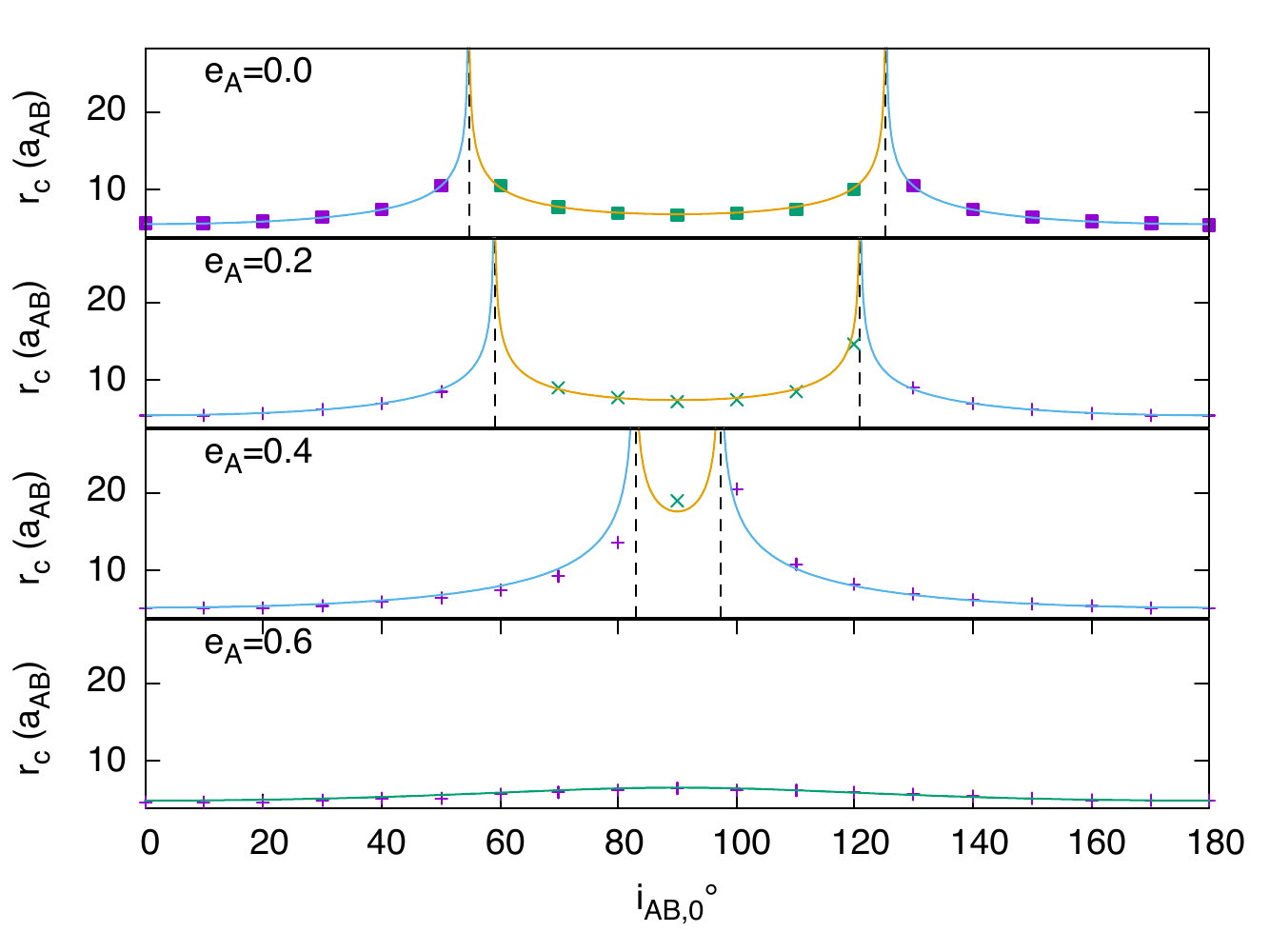}&
\includegraphics[width=.99\columnwidth]{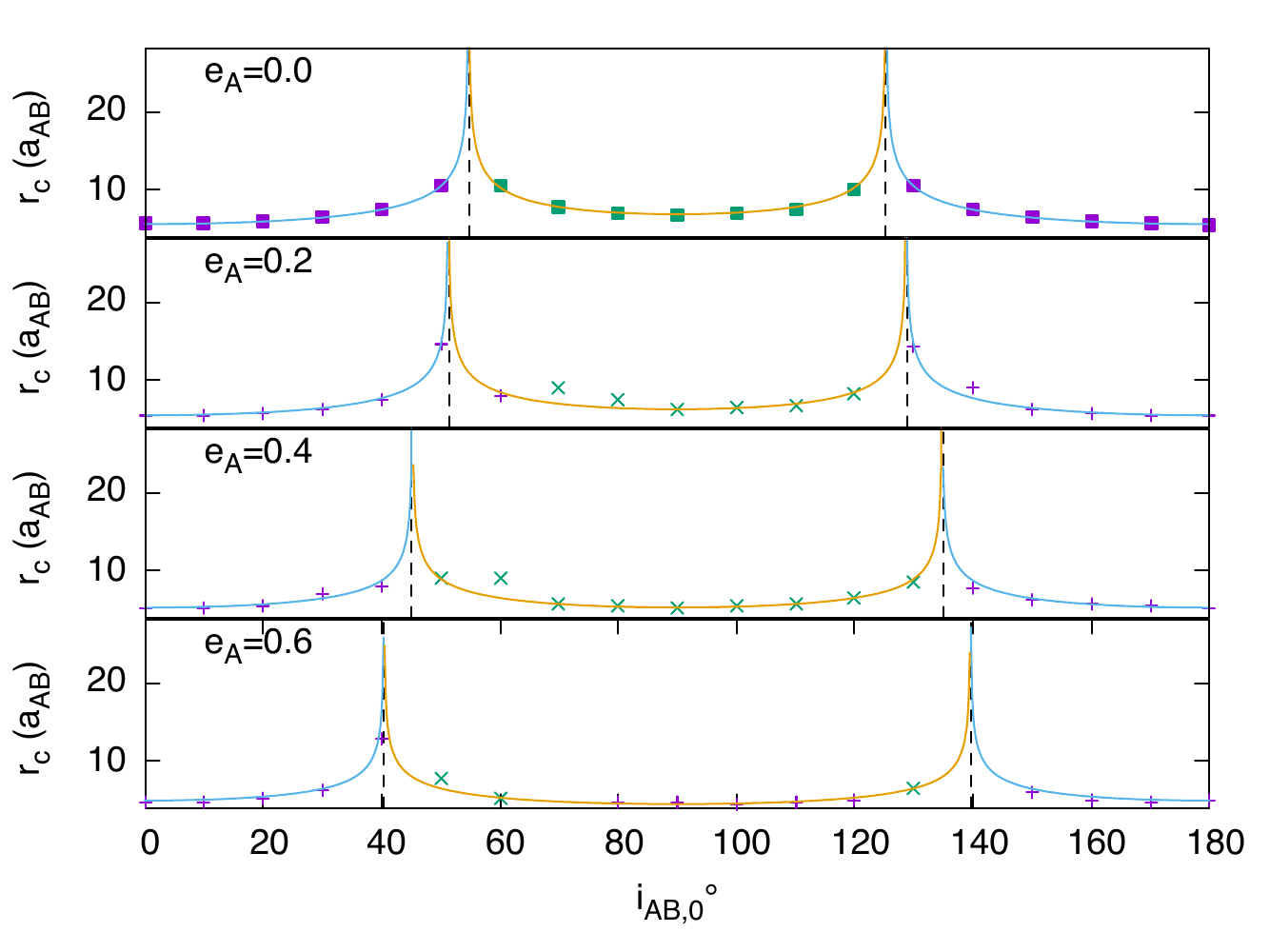}\\
\end{tabular}
\caption{The scaling of the critical radius with $e_A$, in this case the inclination dependence changes and the two angles where the $r_{\rm c}$ goes to infinity move together for Cases A and C and move apart for Cases B and D.  Shown here are Case A on the left and Case B on  the right.}
\label{fig:scale2}
\end{figure*}

Fig.~\ref{fig:scale2} shows how $r_{\rm c}$ is affected by changes to  the initial value of $e_A$. The critical mutual inclination where the apsidal precession rate is zero  depends on $e_A$ (equation~(\ref{eq:ic})). The left panel is case A. 
As $e_A$ increases, the  critical inclinations both move toward $90^\circ$ and when they reach $90^\circ$ at  $e_A\approx 0.408$ there is a broad region near mutual inclination of $90^\circ$ in which the precession rate averages to zero and the circumtriple has stable librating orbits to large radii, as would be the case with a binary. With $e_A>0.408$ the precession rate  never reaches zero and one gets a broad region where the precession is slower than the coplanar case and the critical radius is larger, but there is  no value where it goes to infinity.  The right panel is case B, here the analytical model has the two critical inclinations moving away from each other as $e_{\rm A}$ increases.  

The difference between the cases is the  initial value of $\omega_A$. For case A, $\omega_A=0^\circ$ and the $\cos(2 \omega_A)$ term in the precession rate is $+1$. This is also true for case C. 

For cases B and D, the value of $\omega_A=90^\circ$ and  the $\cos(2 \omega_A)$  term is $-1$.  This term controls whether the precession rate increases or decreases at inclinations near $90^\circ$ for increasing values of $e_A$.  This can be seen in the average precession rate shown in Fig.~\ref{fig:prec2}.  Cases C and D are not shown in Fig.~\ref{fig:scale2} but are very similar to cases A and B, as can be seen in the average precession rate.

Our analytical expression uses the initial value of $\omega_A$ and assumes it doesn't change but in fact $\omega_A$ evolves in our simulations. For low values of the mutual inclination, $i_{AB}$, the value of $\omega_A$ precesses and over the long timesales averages to zero. However, the \SL{ $(1-\cos(i_{AB})^2$)} term in equation~(\ref{eq:rateF}) means the precession rate is not sensitive to $\omega_A$.  The same goes for inclinations near retrograde.  For inclinations near $i_{AB}=90^\circ$, the term is most important. When $\omega_A=90^\circ$ initially, it remains near $\omega_A=90^\circ$ until it goes through a high eccentricity KL cycle and then returns to $\omega_A=90^\circ$ after the cycle is complete.  When $\omega_A=0^\circ$ initially, it  remains near $\omega_A=0^\circ$ until the high eccentricity KL cycle and then returns to $\omega_A=180^\circ$. During further KL cycles it oscillates between these two values.  The $\cos(2 \omega_A)$ term is 1 for both $\omega_A=0^\circ$ and $\omega_A=180^\circ$.  Since the system spends most of the time not in the high eccentrity part of the KL cycle, the values of $\omega_A=0^\circ$ or $\omega_A=90^\circ$ on average work well for our analytical expression. \SL{The cases $\omega_A=0^\circ$ and $\omega_A=90^\circ$ are extrema of the critical inclinations (see Fig.~\ref{fig:icrit}) and so critical inclinations are changing only very slowly near these values.} This behavior is special to these two starting values of $\omega_A$. \SL{We tested a limited number of runs  with other initial values of $\omega_A$. The simulations still produce systems which oscillate around either $\omega_A=90^\circ$ or $\omega_A=270^\circ$ or   a system which transitions between $\omega_A=0^\circ$ and $\omega_A=180^\circ$ but these values occur for different values of $e_A$ and $i_{AB}$ than the initial ones. 
As an  example, with a system starting at our standard setup but with $e_A=0.3$ and $i_{AB}=80^\circ$, then $\omega_A<42^\circ$ and $\Omega_A=0^\circ$ produces an orbit which spends most of its time near $\omega_A=0^\circ$ and $\omega_A=180^\circ$ whereas larger initial values of $\omega_A$ oscillate about  $\omega_A=90^\circ$.  The oscillations are centered about different values of $e_A$ and $i_{AB}$ than the starting values and so our analytical curve using the starting values may not work.   This parameter space is very large and has not been explored in detail.}

\subsection{Triple Star Stability}
Triple star systems are not necessarily stable \citep{Eggleton1996,Livio1998,Mardling2001,Valtonen2006}. In particular at the higher inclinations discussed here are subject to KL oscillations which can lead to very high values of $e_A$ which can cause other processes to become important such as tides \citep{Morais2012,Naoz2016,Hamers2021}.  \cite{Vynatheya2022} have studied the stability of triple star systems and with a machine learning model have produced a python program which predicts the stablity of triple star systems (\cite{Georgakarakos2024} conducted a similar model with planets).  We have explored the region our standard hierarchical system and it is in a region where the program in \cite{Vynatheya2022} predicts stable configurations.  The closest to unstable configurations predicted by  their model from our standard configuration were found by increasing $e_{AB}$ to greater than 0.7 or decreasing $a_{AB}/a_A$ to less than 10.  The studied systems are scale free  but other processes such as general relativity and tides may make some scales unstable and others not.  We haven't considered the long term stability of the systems but just characterize the test particle orbits about them should they exist.

\subsection{No Circulating Region} 

\begin{figure}
\includegraphics[width=.99\columnwidth]{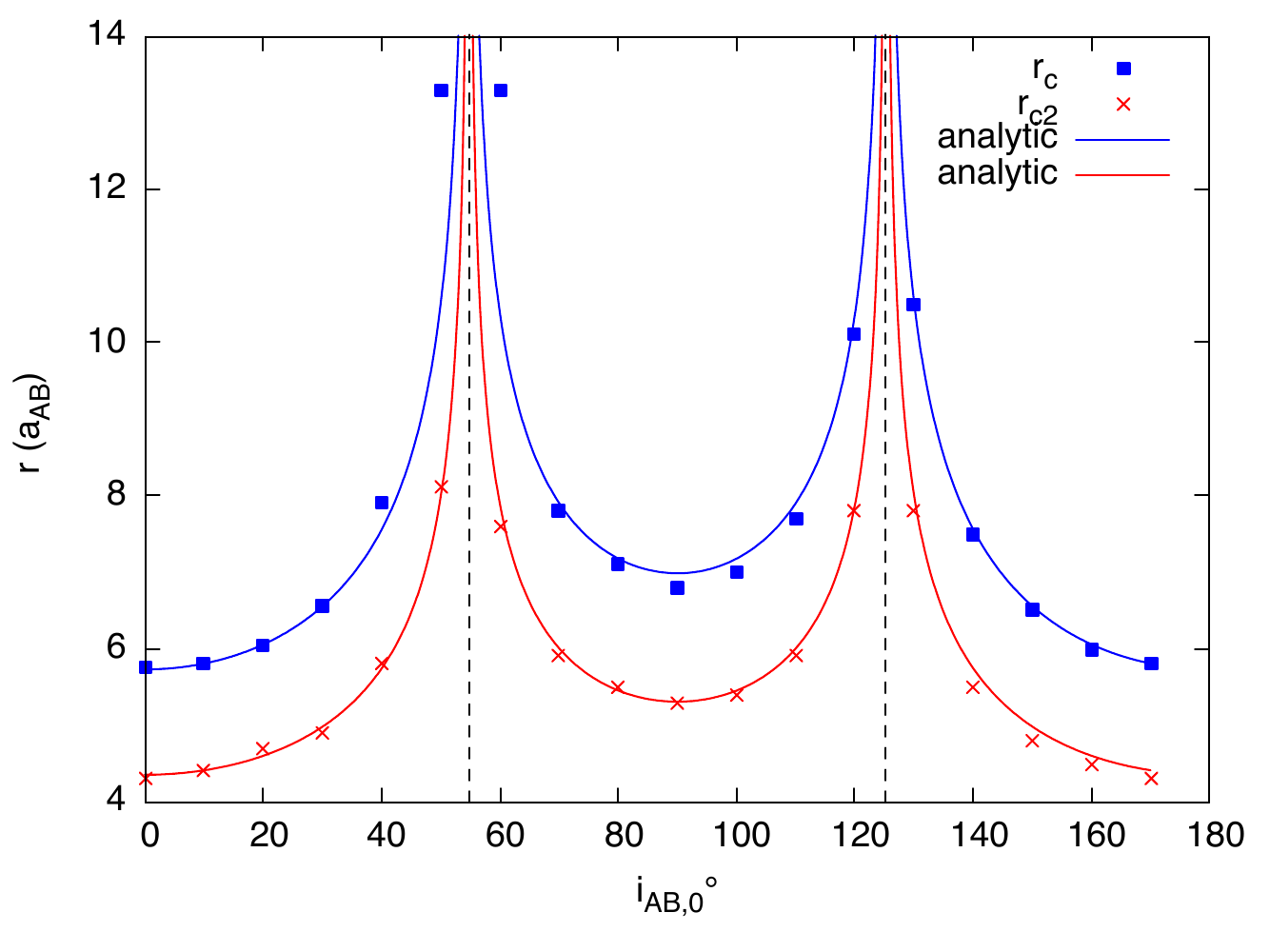}
\caption{For our standard parameters  we plot blue points and curve are the same as Fig.~\ref{fig:maxr} showing critical radius, while the red points and curve show   the radius where the librating region reaches either $180^\circ$ or $0^\circ$.  The analytic curves are equation~(\ref{scaling}).  Between these two radii either retrograde circulating orbits are not possible (for low and high $I_{AB}$) or circulating orbit are impossible between the two $i_c$'s. }
\label{fig:maxl}
\end{figure}

Another interesting radius is the radius where the \SL{inclination of the test particle, $i$,} in the librating region reaches either $180^\circ$ or $0^\circ$ depending on the direction of the precession of the binary. This can be seen in Fig.~\ref{fig:libreg}. When this occurs one of the circulating orbits regions vanishes. 

We label the radius where this occurs $r_{c2}$ and this radius is where the nodal precession rate of the test particle evaluated at either \SL{$i= 180^\circ$ or $i=0^\circ$}  at a phase of $\phi=0^\circ$ is equal to the apsidal precession rate of the outer binary.  This radius is a fixed ratio from the $r_c$ defined above \citep[see][]{Childs2023,Lubow2024} and is given by 
\begin{equation}
r_{c2}=\left(\frac{1-e_{AB}^2}{1+4 e_{AB}^2}\right)^{2/7} r_c\,.   
\label{scaling}
\end{equation}
For our standard model, this radius is shown in Fig.~\ref{fig:maxl} along with the analytical model.  Whether the radius occurs at $i=180^\circ$ or $i=0^\circ$ depends on the direction of apsidal precession, just as $r_c$ does.  For  our standard model with initial inclinations $i_{AB}<55^\circ$ and $i>125^\circ$, the critical radii occur when the librating region reaches $i=180^\circ$ whereas for $55^\circ<i_{AB}< 125^\circ$ it occurs when the libration region reaches $i=0^\circ$.  The critical radius, $r_{c2}$,   is always less than $r_c$ and between these two radii there cannot be a stable circulating orbit, so that for  $55^\circ<i_{AB}< 125^\circ$, there are no stable prograde circulating orbits at radii between $r_{c2}$ and $r_c$ and outside of this mutual inclination range there are no stable retrograde circulating orbits between these two radii.

\section{Discussion and Conclusions}
\label{conc}

We have examined the dynamics of circumtriple test particles around triple stars with large mutual misalignments between the inner and outer binaries with analytic methods and $n$-body simulations. We found good agreement between the two methods.

As a result of apsidal precession of the triple star, there is a critical radius where the nature of the test particle nodal precession changes. For $r<r_{\rm c}$, the particles may be in librating orbits, while for $r>r_{\rm c}$, only circulating orbits occur. We have shown that the critical radius is at a minimum for the coplanar (or retrograde coplanar) circumtriple star system. For larger mutual misalignments, the critical radius is larger. For two values of the mutual misalignment, $i_{\rm c}$ and $180^\circ-i_{\rm c}$, the critical radius becomes infinite since the apsidal precession rate is zero. \SL{In the quadrupole approximation of the potential,} the value of $i_{\rm c}$ is independent of the triple star masses, semi-major axes and outer binary eccentricity. However, it is sensitive to the eccentricity of the inner binary (see equation~\ref{eq:ic}). For a circular orbit inner binary $i_{\rm c}=55^\circ$ and for some configurations (when $\omega_A=0^\circ$ for example) it increases with inner binary eccentricity. For other configurations (when $\omega_A=90^\circ$) it decreases but only slightly and everything in between. For $i_{AB}=i_{\rm c}$, polar librating orbits occur for particles at all orbital radii (as in the circumbinary case).

The motivation for our work was  observations of hierarchical triple stars with mutual misalignments, and observations of circumtriple gas disks that are misaligned to the triple star orbit. Understanding the evolution of test particle orbits is essential for understanding the dynamics of circumtriple disks. However, while the nodal precession of test particle orbits may be periodic on long timescales, a viscous gas disk is subject to dissipation that leads to alignment. The direction of the alignment depends on the type of nodal precession \citep[e.g.][]{Abod2022}. The alignment of the angular momentum vector of the disk is either towards the binary angular momentum vector, or to the stationary inclination. Therefore the different nodal librations have a significant effect on the evolution of gas disks and the alignment of circumtriple planets that may form in such a disk. 

\SL{Our calculations have neglected the effects of general relativity (GR) though we note that it  also provides a small prograde precession \citep{Zanardi2018,Lepp2022}.  Including GR would break our scaling relations, but the precession from GR for most stellar systems would be much smaller than the precession from the three body interactions discussed here.  For example, in our standard system the GR alone would have $r_c\approx 157 a_{AB}$  using equation 7 from \cite{Lepp2022}. This is due to the precession from GR on the binary but the difference to $r_c$ would be too small to show in our figures.  For a near stellar mass inner binary it is likely also to be small but for more massive black hole binary systems the GR precession on the inner binary could significantly affect the radius for $r_{\rm c}$ \citep{Childs2024,Martin2024}. In addition, GR can induce loss of energy to gravitational waves and speed the merger of the inner binary  \citep{Antonini2012,Lepp2023b}.}

In conclusion, we suggest that polar circumtriple disks may be more likely around triple star systems with large mutual misalignments compared to coplanar (or retrograde coplanar) systems.  Specifically, if the  triple star has mutual mistalignment $i_{\rm c}$ or $180^\circ-i_{\rm c}$, then polar alignment can occur no matter the radial extent of the disk.   While it is beyond the scope of this current work, this should be investigated in the future with hydrodynamical disk simulations. 

\section*{Acknowledgements}
We would like to thank the referee for useful comments.  We acknowledge support from NASA through grants 80NSSC21K0395, 80NSSC19K0443 and 80NSSC23M0104.

\bibliographystyle{aasjournal}
\bibliography{ct} 

\end{document}